\documentclass[10pt,fullpage]{article}

\usepackage{fullpage}
\usepackage{graphicx}
\usepackage{amsmath}
\usepackage{amssymb}
\usepackage{amstext}
\usepackage{amsthm}
\usepackage{epsfig}
\usepackage{placeins}
\usepackage{subfig}
\usepackage{paralist}
\usepackage{caption}
\usepackage{multirow}
\usepackage{algorithm}
\usepackage{algorithmic}
\usepackage{comment}
\usepackage[obeyspaces,spaces]{url}
\usepackage{balance}
\usepackage{enumitem}
\usepackage{hyperref}
\usepackage{times}

\makeatletter
 {\vskip 5pt\begin{list}{}{%
  \setlength{\topsep}{0pt}\setlength{\partopsep}{0pt}\setlength{\parskip}{0pt}%
  \setlength{\parsep}{0pt}\setlength{\labelwidth}{0pt}%
  \setlength{\rightmargin}{0pt}\setlength{\leftmargin}{0pt}%
  \setlength{\labelsep}{0pt}%
  \obeylines\@vobeyspaces\normalfont\ttfamily%
  \item[]}}
 {\end{list}\vskip5pt\noindent}
\makeatother

\begin{document}

\date{March 2018}

\title{Distributed Caching for Complex Querying of Raw Arrays}

\author{Weijie Zhao$^{1}$, Florin Rusu$^{1,2}$, Bin Dong$^{2}$, Kesheng Wu$^{2}$, Anna Y. Q. Ho$^{3}$, and Peter Nugent$^{2}$\\
\{wzhao23,frusu\}@ucmerced.edu, \{dbin,kwu\}@lbl.gov, ah@astro.caltech.edu, penugent@lbl.gov\\
$^{1}$University of California Merced\\
$^{2}$Lawrence Berkeley National Laboratory\\
$^{3}$Cahill Center for Astrophysics, California Institute of Technology
}

\maketitle

\begin{abstract}

As applications continue to generate multi-dimensional data at exponentially increasing rates, fast analytics to extract meaningful results is becoming extremely important. The database community has developed array databases that alleviate this problem through a series of techniques. In-situ mechanisms provide direct access to raw data in the original format---without loading and partitioning. Parallel processing scales to the largest datasets. In-memory caching reduces latency when the same data are accessed across a workload of queries. However, we are not aware of any work on distributed caching of multi-dimensional raw arrays.
In this paper, we introduce a distributed framework for cost-based caching of multi-dimensional arrays in native format. Given a set of files that contain portions of an array and an online query workload, the framework computes an effective caching plan in two stages. First, the plan identifies the cells to be cached locally from each of the input files by continuously refining an evolving R-tree index. In the second stage, an optimal assignment of cells to nodes that collocates dependent cells in order to minimize the overall data transfer is determined. We design cache eviction and placement heuristic algorithms that consider the historical query workload. A thorough experimental evaluation over two real datasets in three file formats confirms the superiority -- by as much as two orders of magnitude -- of the proposed framework over existing techniques in terms of cache overhead and workload execution time.

\end{abstract}

\section{INTRODUCTION}\label{sec:intro}

In the era of big data, many scientific applications -- from high-energy physics experiments to cosmology telescope observations -- collect and analyze immense amounts of data at an unprecedented scale. For example, projects in astronomy such as Sloan Digital Sky Survey\footnote{\url{http://www.sdss.org/dr13/}} (SDSS) and Palomar Transient Factory\footnote{\url{http://www.ptf.caltech.edu/iptf/}} (PTF) collect the observations of stars and galaxies at a nightly rate of hundreds of gigabytes. The newly-built Zwicky Transient Facility\footnote{\url{https://www.ptf.caltech.edu/ztf}} (ZTF) records the sky at a 15X higher rate than PTF---up to 7 TB per night. A common characteristic of the datasets produced by astronomy projects is that the data are natively organized in a \textit{multi-dimensional array} rather than an unordered set. Due to the inefficacy of traditional relational databases to handle ordered array data~\cite{scidb,ssdb}, a series of specialized array processing systems~\cite{rasdaman,ram,scidb,SciQL-ideas,scihadoop,extascid} have emerged. These systems implement natively a distributed multi-dimensional array data model in which arrays are chunked across a \textit{distributed shared-nothing cluster} and processed concurrently.

\textbf{Automatic transient classification.}
The PTF project aims to identify and automatically classify transient astrophysical objects such as variable stars and supernovae in real-time. A list of potential transients -- or candidates -- is extracted from the images taken by the telescope during each night. They are stored as a sparse array:\\
\hspace*{1cm}\texttt{candidates<bright,mag,$\dots$>[ra,dec,time]}\\
with three dimensions -- the equatorial coordinates \texttt{ra} and \texttt{dec}, and \texttt{time} -- and tens to a few hundred attributes such as brightness and magnitude. The range of each dimension is set according to the accuracy of the telescope which increases with every generation of lenses. In PTF~\cite{ptf:overview}, an array cell corresponds to 1 arcsecond on \texttt{ra} and \texttt{dec}, and 1 minute on \texttt{time}, which generate more than $10^{15}$ cells. The \texttt{candidates} array is stored in several FITS\footnote{\url{https://fits.gsfc.nasa.gov/fits_documentation.html}} files---one for every night. Since the telescope is pointed to distinct parts of the sky each night, the files cover different ranges of the array space. However, the ranges are large enough to overlap. The number of candidates in a file has high variance---there are sparse files with only tens of candidates and skewed files with millions of candidates.

Transient classification consists of a series of array queries that compute the similarity join~\cite{Zhao:array-sim-join} between candidates, i.e., find all the candidates identified at similar coordinates, possibly in a time window. In order to execute these queries inside an array database, the \texttt{candidates} array has to be loaded and partitioned. This process is time-consuming -- it takes hours in SciDB according to~\cite{extascid,Xing-Arraybridge} -- and duplicates the data. As a result, it is infeasible for the real-time processing required in PTF. In-situ processing over raw files~\cite{nodb,scanraw,sdsq} is an alternative that eliminates loading altogether and provides instant access to data. While there is a large body of work on in-situ processing for CSV and semi-structured files such as JSON and Parquet~\cite{nodb,sinew,scanraw,Olma-Slalom}, there are only two extensions to multi-dimensional arrays~\cite{Xing-Arraybridge,Han-distributed-in-situ-array}. Both of them act as optimized HDF5\footnote{\url{https://support.hdfgroup.org/HDF5/}} connectors for SciDB. They allow SciDB to execute declarative array queries over dense arrays stored in HDF5 format by pushing down the subarray operator into the HDF5 read function. None of the connectors support the sparse and skewed files corresponding to PTF \texttt{candidates} because these files are not organized and partitioned along the dimensions of the array. Even for dense arrays, the initial partitioning specified at array definition may be suboptimal with respect to the query workload. For example, a tight band range on one dimension triggers a high number of partition accesses even though the number of queried cells is small. Moreover, the connectors do not provide any caching mechanism. This is pushed entirely to the SciDB buffer manger which operates at instance-level, i.e., it caches only local arrays. This precludes any optimization in accessing the raw data because the granularity and the content to cache are query-dependent rather than partition-dependent---partition or chunk caching is suboptimal. As pointed out in~\cite{nodb,Ailamaki:proteus,Olma-Slalom}, caching is extremely important for in-situ processing because repeated access to raw data is expensive---the case for the set of queries in PTF transient classification.

\textbf{Problem statement.}
In this paper, we tackle the problem of \textit{distributed caching for raw arrays} with the goal to accelerate queries over frequently accessed ranges. Given a set of raw files that contain portions of an array and an online dynamic query workload, we have to determine which cells to cache in the distributed memory of an array database system. Rather than having each node manage its memory buffer to cache local cells, we aim for a global caching infrastructure that automatically identifies both the cells to cache and the instance where to cache them. Our ultimate goal is to provide both instant access to distributed multi-dimensional raw arrays and optimal query performance through caching.

\textbf{Challenges.}
Distributed caching for arrays poses two main challenges. The first challenge is to determine the cells kept in the cache at query time. This is a problem because cache misses over raw arrays are very expensive. Even if we have only one cache miss for a cell, it requires us to scan all the raw files whose bounding box contains the cell---for dense arrays, although direct access eliminates scanning, a complete disk read is still necessary. However, at most one file contains the required cell. The second challenge comes from the distributed nature of array database. The conventional approach caches the requested data at the origin instance---where it is stored. This approach works well in content delivery networks (CDN)~\cite{Ousterhout-RAMCloud,cdn-caching} where data co-locality is not required for query execution. However, most of the array queries specify shape-based relationships between cells in the form of stencil operators and their generalization to similarity joins~\cite{Zhao:array-sim-join}. Given the current placement of raw and cached data, the challenge is to coordinate and organize the caches across nodes -- decide what cached data are placed on which nodes -- in order to preserve data co-locality and provide efficient data access. Direct application to arrays of generic distributed memory caching~\cite{Floratou-AdaptiveCaching,Li-Tachyon,GRAPPA,Altinel-CacheTables,wang2007workload} implemented in Hadoop and Spark suffers from excessive communication and load imbalance due to the skewed distribution of cells across files.

\textbf{Approach.}
We design a distributed caching framework that computes an effective caching plan in two stages. First, the plan identifies the cells to be cached locally from each of the input files by continuously refining an evolving R-tree index. Each query range generates a finer granularity bounding box that allows advanced pruning of the raw files that require inspection. This guarantees that -- after a sufficiently large number of queries -- only relevant files are scanned. In the second stage, an optimal assignment of cells to nodes that collocates dependent cells in order to minimize the overall data transfer is determined. We model cache eviction and placement as cost-based heuristics that generate an effective cache eviction plan and reorganize the cached data based on a window of historical queries. We design efficient algorithms for each of these stages. In the long run, the reorganization improves cache data co-locality by grouping relevant portions of the array and by balancing the computation across nodes.

\textbf{Contributions.}
The specific contributions of this paper can be summarized as follows:
\begin{itemize}[leftmargin=*,noitemsep,nolistsep]
\item We introduce distributed caching for raw arrays (Section~\ref{sec:distributed-caching}). While caching over raw data has been explored in a centralized setting, this is the first work that investigates in-situ processing with distributed caching.
\item We design an evolving R-tree index that refines the chunking of a sparse array to efficiently find the cells contained in a given subarray query (Section~\ref{ssec:distributed-caching:raw-file}). The index is used to eliminate unnecessary raw files from processing.
\item We propose efficient cost-based algorithms for distributed cache eviction and placement that consider a historical query workload (Section~\ref{ssec:distributed-caching:replacement} and~\ref{ssec:distributed-caching:placement}). The goal is to collocate dependent cells in order to minimize data transfer and balance computation.
\item We evaluate experimentally our distributed caching framework over two real sparse arrays with more than 1 billion cells stored in three file formats---CSV, FITS, and HDF5 (Section~\ref{sec:experiments}). The results prove the effectiveness of the caching plan in reducing access to the raw files and the benefits of the eviction algorithm. 
\end{itemize}

\noindent
While our solution is presented for sparse arrays, the proposed framework is also applicable to dense data. While access to a specific cell is much faster in a dense array, caching entire chunks incurs unnecessary memory usage. Instead of caching the complete chunks as specified in the array definition, our solution infers a workload-derived caching scheme that may have smaller granularity. This allows for more relevant data to be cached. Another important application of this work is to distributed linear algebra which is the fundamental building block in large-scale machine learning (ML). Stochastic gradient descent (SGD)~\cite{sgd-gpu} -- the primary algorithm to train ML models -- requires access to rows in the example matrix that have non-zero entries for certain columns---or features. Existing solutions in Spark MLlib~\cite{mllib} and TensorFlow~\cite{tensorflow} access all the blocks containing the columns of the required example---which is unnecessary and expensive. Our framework builds an evolving R-tree index that allows for non-zero columns to be easily found without accessing the block.

\section{PRELIMINARIES}\label{sec:preliminaries}

In this section, we introduce multi-dimensional arrays, in-situ data processing, and distributed caching---the foundation for distributed caching of raw arrays.

\begin{figure}[htbp]
\begin{center}
 \includegraphics[width=.5\textwidth]{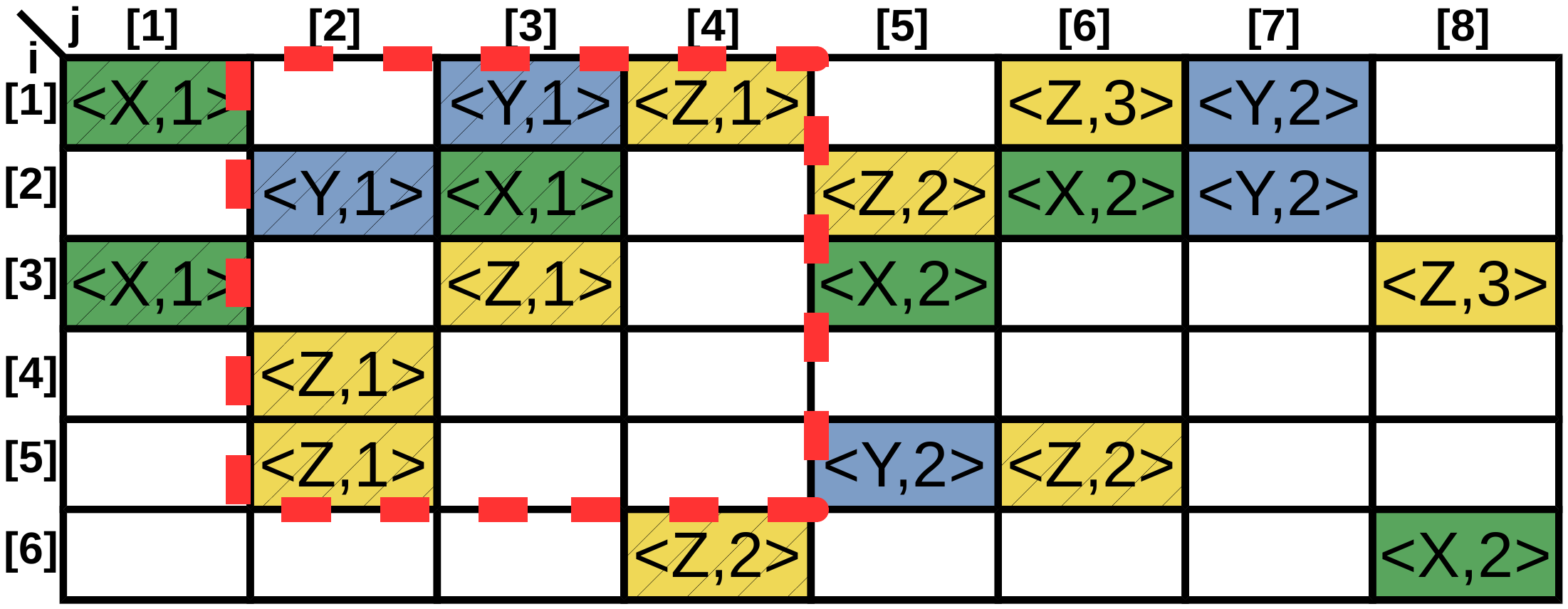}
\end{center}
\caption{Array \texttt{A<n:char,f:int>[i=1,6;j=1,8]}. This is a raw array consisting of 7 files distributed over 3 nodes. Attributes \texttt{n} and \texttt{f} correspond to the node and the file id on the node, e.g., \texttt{<X,2>} is a cell in the second file on node X. The cell color also shows the server on which the file is stored. The dashed rectangle specifies a query.}
\label{fig:array-example}
\end{figure}

\subsection{Multi-Dimensional Arrays}\label{ssec:prelims:ADM}

A multi-dimensional array is defined by a set of \textit{dimensions} $\mathcal{D}=\{D_{1}, D_{2}, \dots, D_{d}\}$ and a set of \textit{attributes} $\mathcal{A}=\{A_{1}, A_{2}, \dots, A_{m}\}$. Each dimension $D_{i}$ is a finite ordered set. We assume that $D_{i}, i\in[1,d]$ is represented by the continuous range of integer values $[1,N]$. Each combination of dimension values, or indices, $\left[i_{1}, i_{2}, \dots, i_{d}\right]$, defines a \textit{cell}. Cells have the same scalar type, given by the set of attributes $\mathcal{A}$. Dimensions and attributes define the schema of the array. Based on these concepts, an array is defined as a function over dimensions and taking value attribute tuples:
\begin{equation*}
	\textit{Array} : [D_{1},D_{2},\dots,D_{d}] \longmapsto \left<A_{1}, A_{2}, \dots, A_{m}\right>
\end{equation*}
A 2-D array \texttt{A<n,f>[i=1,6;j=1,8]} with dimensions \texttt{i} and \texttt{j} and attributes \texttt{n} and \texttt{f} is depicted in Figure~\ref{fig:array-example}. This is the notation to define arrays in SciDB's AQL language~\cite{aql:algebra,aql:syntax}. The range of \texttt{i} is $[1,6]$, while the range of \texttt{j} is $[1,8]$. Non-empty cells are colored.

\textbf{Chunking.}
Array databases apply chunking for storing, processing, and communicating arrays. A \textit{chunk} groups adjacent array cells into a single access and processing unit. While many strategies have been proposed in the literature -- see~\cite{array-survey} for a survey -- regular chunking is the most popular in practice, e.g., SciDB. Regular chunks have the same dimensionality as the input array, are aligned with the dimensions, and have the same shape. The cells inside a chunk are sorted one dimension at a time, where dimensions are ordered according to the array schema. In the case of sparse and skewed arrays, regular chunking produces imbalanced chunks. Moreover, if the input data is not fully known, it is impossible to choose an appropriate chunking scheme. In \textit{dynamic workload-based chunking}~\cite{furtado:tiling,Heinis:thermal-join}, the array is continuously partitioned based on the incoming queries and only for the ranges accessed in the query. This strategy is the most adequate for raw sparse arrays since no information on the content is available.

\textbf{Shared-nothing architecture.}
We are given a distributed array database having a \textit{shared-nothing architecture} over a cluster of $N$ \textit{servers} or \textit{nodes}, each hosting one instance of the query processing engine and having its local attached storage. The chunks of each array are stored across several servers in order to support parallel processing. All servers participate in query execution and share access to a centralized system catalog that maintains information about active servers, array schemas, and chunk distribution. A \textit{coordinator} stores the system catalog and manages the nodes and their access to the catalog. The coordinator is the single query input point into the system. For example, in Figure~\ref{fig:array-example} there are $3$ servers in the database. The chunks of array \texttt{A} -- which are complete files in this case -- are distributed over the $3$ servers. The color of the cell corresponds to the assigned server, while the attributes identify the file the cell belongs to, e.g., \texttt{<X,2>} is a cell in file $2$ of server X.

\subsection{Array Similarity Join}\label{ssec:prelims:array-sim-join}

Array similarity join is introduced in~\cite{Zhao:array-sim-join} as a generalization of array equi-join~\cite{Duggan:array-joins} and distance-based similarity join. This operator can encode any join conditions on dimensions and it can express the most common similarity distances, e.g., $L^{p}$ norms, Earth Mover's Distance (EMD), and Hamming~\cite{Zhao:array-sim-join}. For example, the $L^{1}(1)$ similarity self-join of the dashed subarray in Figure~\ref{fig:array-example} is a 4-D array with $17$ non-empty cells---there is a cell for each non-empty cell in the subarray and 10 additional cells are generated by pairs of adjacent cells that have $L^{1}$ distance of 1. The execution of the array similarity join operator requires identifying the chunks that have to be joined and the assignment of these pairs to nodes that perform the join. An optimization algorithm that minimizes the overall query processing time and the amount of transferred data is given in~\cite{Zhao:array-sim-join}. The optimal plan is computed at the coordinator from the catalog metadata which stores the location of the chunks and relevant chunk statistics. It consists of a detailed chunk transfer and execution sub-plan for each node. This approach is applicable to any distributed join processing algorithm over range-based partitioned data---only the number of chunk pairs varies.

\subsection{In-Situ Processing over Raw Data}\label{ssec:prelims:raw}

Multiple systems that perform queries over raw data that do not reside in databases have been proposed~\cite{files-queries-results,nodb,instant-loading,data-vaults,scanraw,invisible-loading,Ailamaki:proteus} recently. They are extensions of the external table mechanism supported by standard database servers. These systems execute \texttt{SQL} queries directly over raw data while optimizing the conversion process into the format required by the query engine. This eliminates loading and provides instant access to data. Several systems~\cite{nodb,invisible-loading,scanraw} provide a dynamic tradeoff between the time to access data and the query execution time by adaptively loading a portion of the data during processing. This allows for gradually improved query execution times while reducing the amount of storage for replication. In-situ processing over scientific data -- specifically, multi-dimensional arrays -- resumes to connectors from a query processing engine, e.g., SciDB, to an I/O-library for a specific data format, e.g., HDF5~\cite{sdsq,Xing-Arraybridge,Han-distributed-in-situ-array}. These systems allow SciDB to execute declarative array queries exclusively over dense HDF5 arrays by pushing-down the subarray operator. They do not provide integrated optimizations across layers. The focus of this work is on the more general case of unorganized (sparse) arrays---not only HDF5.

\subsection{Caching}\label{ssec:prelims:caching}

All database systems cache disk pages in memory buffers after they are read from disk. The fundamental problem in caching is to decide which pages to keep in memory when the available budget is exhausted---or, equivalently, which pages to evict. The provably optimal eviction algorithm is to remove the page that will be accessed farthest in the future. Since the future is generally unknown, existing eviction algorithms are based on the past query workload. The least-recently-used (LRU) and least-frequently-used (LFU) are the most common eviction algorithms widely implemented in computer systems. They can be easily extended to array databases by replacing the page with the chunk~\cite{scidb}. Online cost-based caching algorithms~\cite{cost-cache-algo,cost-cache-web} prioritize the eviction of pages/chunks with lower cost in order to keep the expensive items in cache, where the cost is application-dependent. Since there are no provable polynomial time cost-based cache eviction algorithms~\cite{cost-cache-algo}, Greedy heuristics are the only solution. Nonetheless, these algorithms are preferred in a distributed setting where the cost of accessing data is variable. The standard approach in distributed caching is to define a global shared memory that is mapped to the local memory of physical nodes~\cite{Li-Tachyon,GRAPPA}. This allows for an immediate extension of the centralized algorithms. Parallel databases use a simpler approach in which each node is managing its local cache---there is no global cache manager. The reason is transactional cache consistency. However, in read-intensive scientific applications, transactions are not an issue. Caching for raw data processing is necessary because of the high cost to extract data. The NoDB~\cite{nodb} and Proteus~\cite{Ailamaki:proteus} systems cache everything in memory and use LRU for eviction, while SCANRAW~\cite{vert-part-offline} materializes the evicted pages adaptively in the database storage. However, they do not target raw sparse arrays and are centralized solutions.

\section{Raw Array Distributed Caching}\label{sec:distributed-caching}

In this section, we present the first distributed caching mechanism over raw array data proposed in the literature. We begin with a high-level description of the distributed caching inner-workings that identifies the main processing stages. Then, we delve into the details of each stage and introduce our technical contributions.

\begin{figure}[ht]
\centering
	\includegraphics[width=0.5\textwidth]{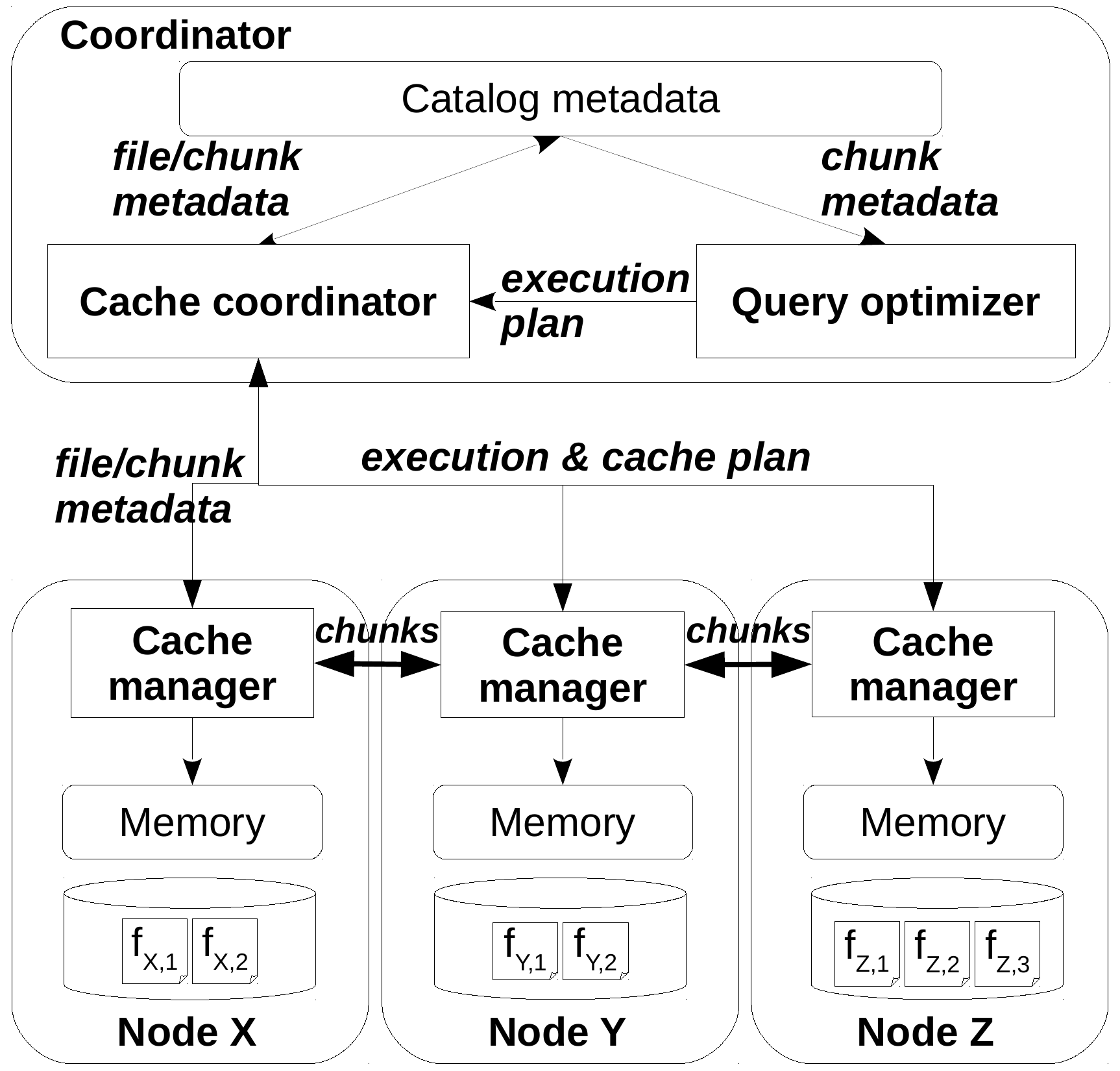}
	\caption{Architecture of distributed caching for raw arrays.}
	\label{fig:architecture}
\end{figure}

\textbf{Problem setting.}
Given a collection of raw files distributed across the nodes of the array database and a query workload, the goal is to process the queries optimally with limited memory budget for caching. We assume the entire workload is unknown before processing---queries are processed one after another in online fashion. Figure~\ref{fig:architecture} illustrates the proposed distributed caching architecture which consists of a cache coordinator and a cache module for every node of the array database. These can run either as independent processes or as threads that are part of the array database coordinator and nodes, respectively. Each file $f_{i,j}$ -- the $j^{\text{th}}$ file stored in its entirety on node $i$ -- has a bounding box $\mathcal{B}(f_{i,j})$ -- stored in the catalog at the cache coordinator -- that contains the range of each dimension for the cells in the file. The cache coordinator has a complete view of the distributed cache memory across nodes both in terms of content and size. This allows for optimizing in-memory chunk location, thus, query execution, and is a major departure from distributed buffer managers such as Spark's. We emphasize that the cache coordinator does not receive any file data---exclusively managed by the cache nodes.

\textbf{Approach.}
The array distributed caching framework works as follows (Figure~\ref{fig:architecture}). First, the execution engine sends the subarray and the shape in the array similarity join query to the cache coordinator which has to determine the chunk pairs that have to be joined. While these can be immediately inferred from the catalog for loaded data, there is no chunking -- or the chunking is not known -- for raw data. The naive solution to handle this problem is to treat each file as a chunk and perform the similarity join between pairs of files. Since a file covers a much larger range and is not organized based on cell locality, this solution accesses many unnecessary cells and uses the cache budget inefficiently. Our approach is to build in-memory chunks incrementally based on the query---workload-driven chunking~\cite{array-survey}. These chunks are built by the cache nodes by accessing the files that overlap with the query---the only data the cache coordinator forwards to the nodes. Instead of extracting chunks only for the queried data, the cache nodes create higher granularity chunks for the entire file. Since chunk building requires raw file access -- which is expensive -- caching is very important for efficient processing. In the best case, the entire file can be cached. Otherwise, a cache replacement policy is required to determine the chunks to keep in memory. Once evicted from memory, the chunks are lost---they have to be recreated from the raw data. While it is possible to materialize the evicted chunks in some representation, e.g., R-tree, this replicates data and adds another dimension -- which representation to read data from -- to the problem.

Instead of allowing each node to execute a local cache replacement algorithm, e.g., LRU, we develop a global cost-based caching mechanism~\cite{cost-based-caching} executed at the coordinator. This algorithm takes the metadata of all the created chunks across nodes and determines the chunks to be evicted from the cache in a unified way---this may incur chunk transfer across nodes. Since query evaluation also requires chunk transfer and replication, we go one step further and combine query processing with cache replacement. Specifically, cache replacement piggybacks on the query execution plan and eliminates extra chunk transfer. Moreover, the location of a cached chunk is computed by considering the relationship with other chunks in the historical query workload. This is implemented as follows (Figure~\ref{fig:architecture}). Upon receiving the chunk metadata from the nodes, the coordinator -- query optimizer -- computes the optimal query execution plan which specifies what node joins each of the chunk pairs. This is passed to the cache coordinator which computes the cache replacement plan by combining the execution plan with chunk usage statistics and chunk co-locality information. The cache replacement plan informs every node on what chunk replica to cache and which chunks to evict. Finally, these two plans are forwarded to the nodes for execution---the engine executes the query and the cache executes the replacement plan, in this sequence.

\subsection{Raw Array Chunking}\label{ssec:distributed-caching:raw-file}

The goal of raw array chunking is to infer the chunks of the array file dynamically at runtime from the query workload. Instead of creating arbitrary chunks during loading -- which is time-consuming, delays the time-to-query, and may not be optimal for the actual workload -- we build chunks incrementally one query-at-a-time. Given a subarray in the domain of the raw arrays over which to evaluate a similarity join query, we have to identify the cells that are relevant for the query while minimizing the total number of inspected cells. We have to avoid access to the raw files because they are unorganized and require full scan. This cannot be realized initially unless there is no overlap between a file and the query. A straightforward solution is to load the coordinates of all the cells after the first access. However, this is impractical because it requires complete data replication. Instead, we build chunks that group close cells and maintain only the much smaller bounding box. The chunks are cached in the memory of the nodes, while the bounding boxes are managed by the cache coordinator in memory. Moreover, not all the chunks can be cached.

\begin{figure*}[htbp]
\begin{center}
\subfloat[]{\includegraphics[width=.09\textwidth]{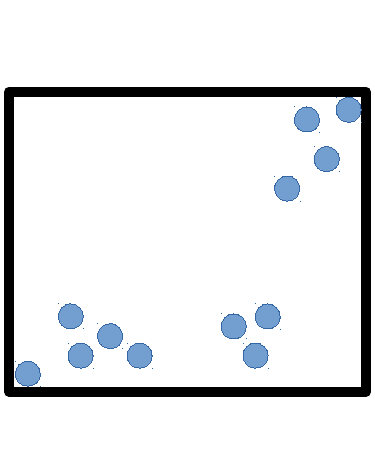}}
\hspace*{0.2cm}
\subfloat[]{\includegraphics[width=.24\textwidth]{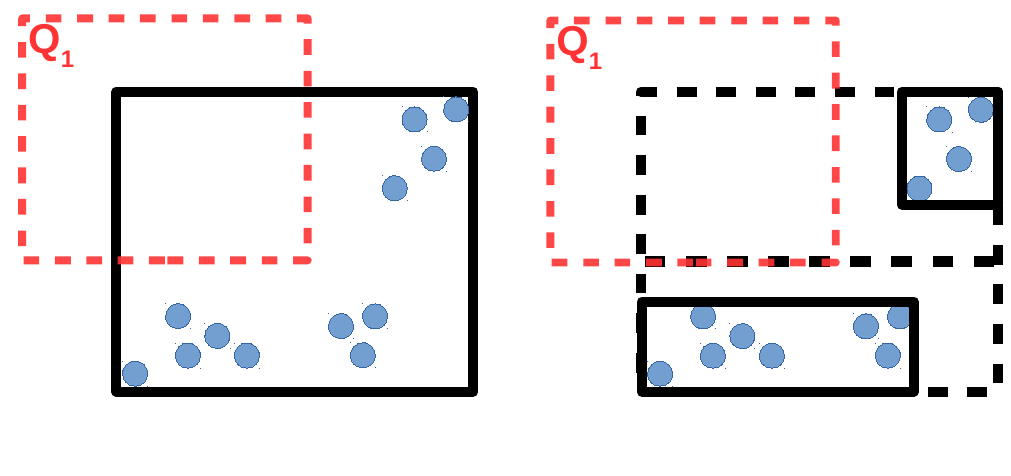}}
\hspace*{0.2cm}
\subfloat[]{\includegraphics[width=.24\textwidth]{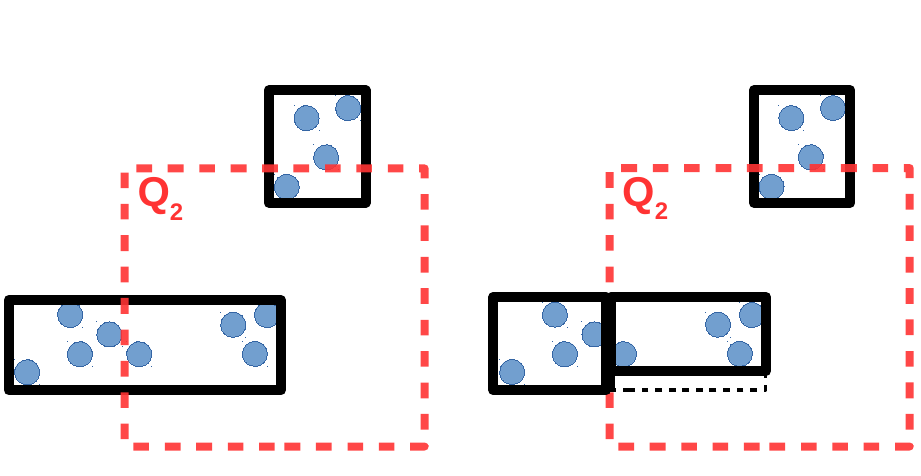}}
\hspace*{0.2cm}
\subfloat[]{\includegraphics[width=.2\textwidth]{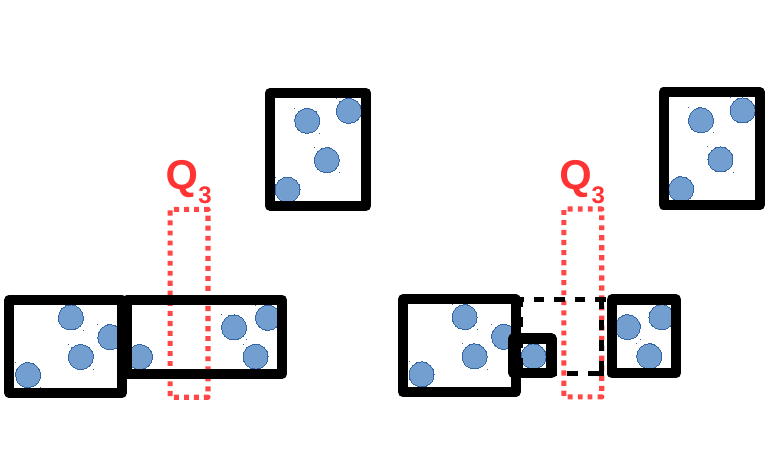}}
\hspace*{0.2cm}
\subfloat[]{\includegraphics[width=.1\textwidth]{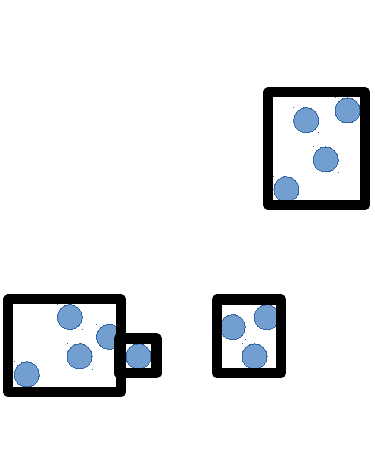}}
\end{center}
\caption{Raw array query-driven chunking. (a) depicts a raw array and its bounding box. (b), (c), and (d) show how the chunking evolves based on a workload of three queries. (e) depicts the resulting four chunks into which the initial array is partitioned.}
\label{fig:raw-chunking}
\end{figure*}

We design a novel incremental chunking algorithm that builds an evolving R-tree~\cite{rtree} based on the queries executed by the system. The invariant of this algorithm is that the set of chunks cover all the cells of the array at any time instant. Moreover, the chunks are non-overlapping. The central point of this algorithm is splitting a chunk that overlaps with the query subarray (Algorithm~\ref{alg:evolving}). Two questions require answer.

\noindent
\textbf{When to split?} A chunk is split in two cases. First, if there are a sufficiently large number of cells in the chunk. This threshold is a configurable parameter that can be set by the user. In the second case, even when the number of cells is below the threshold, if the query subarray does not contain any cell, the chunk is split further. Creating a large number of (small) chunks has both positive and negative impact. On the positive side, the likelihood that a chunk that overlaps with the query contains relevant cells is higher. This avoids inspecting unnecessary chunks. On the negative side, the number of bounding boxes that have to be managed by the cache coordinator increases and this makes query and cache optimization more time-consuming.

\noindent
\textbf{How to split?} The number of additional chunks generated by a split varies between $1$ and $3^{d}$, where $d$ is the dimensionality of the array. We opt for always splitting a chunk into two chunks. This is done by selecting a single splitting dimension. The algorithm enumerates over the queried subarray boundaries that intersect with the chunk bounding box and chooses to split into those two chunks that have the minimum combined hyper-volume. Rather than computing the hyper-volume from the query-generated chunks, we derive the bounding box of a chunk only from the cells assigned to it. Typically, this results in smaller and more condensed chunks. We increase the number of chunks conservatively because we do not want the number of bounding boxes managed by the cache coordinator to explode with the number of queries. Chunks that cover a smaller hyper-volume are more compact, thus the probability to contain relevant cells is higher.

\begin{algorithm}[htbp]
\caption{Chunk Split}\label{alg:evolving}

\algsetup{linenodelimiter=.}

\begin{algorithmic}[1]

\REQUIRE Chunk $\alpha$ with bounding box $\mathcal{BB}_{\alpha}$ that intersects query subarray $Q$; Minimum number of cells threshold $\textit{MinC}$

\ENSURE Chunks $\beta$ and $\gamma$ after splitting $\alpha$

\STATE \textbf{if}(cells in $\alpha$ < $\textit{MinC}$) \AND ($\exists$ cell in $\alpha$ $\in$ $Q$) \textbf{then} \textbf{return}

\STATE min\_vol = $+\infty$
\FOR{\textbf{each} boundary $b \in Q$ that intersects with $\mathcal{BB}_{\alpha}$}
	\STATE $(\beta_{b},\gamma_{b})$ $\leftarrow$ split cells in $\alpha$ into two sets by boundary $b$
	\IF{vol($\beta_{b}$) + vol($\gamma_{b}$) < min\_vol}
		\STATE min\_vol $\leftarrow$ vol($\beta_{b}$) + vol($\gamma_{b}$)
		\STATE $\beta \leftarrow$ bounding\_box($\beta_{b}$), $\gamma \leftarrow$ bounding\_box($\gamma_{b}$)
	\ENDIF
\ENDFOR
\end{algorithmic}

\end{algorithm}

We illustrate how the algorithm works for a 2-D array in Figure~\ref{fig:raw-chunking}. The dashed box is the query subarray $Q$, while the set of chunks at a given instant is represented by solid bounding boxes. Initially, there is a single chunk with a large bounding box. This corresponds to the root of the R-tree. Since only two boundaries of query $Q_{1}$ overlap with the original chunk, only two potential splits are considered. The one that is chosen corresponds to the horizontal axis. Two smaller chunks are generated. Query $Q_{2}$ overlaps with both of these two chunks. Since the number of cells in the upper chunk is $4$ -- smaller than the splitting threshold of $5$ -- and there is a cell relevant to the query, no split is triggered. The lower chunk is split based on the single query boundary that intersects with it. Query $Q_{3}$ overlaps only with one of the input $3$ chunks. Although the number of cells in the chunk is below the splitting threshold, there is no cell in the query subarray. This triggers a split into two condensed chunks, for a total of four chunks overall.

\begin{figure*}[htbp]
\begin{center}
 \includegraphics[width=\textwidth]{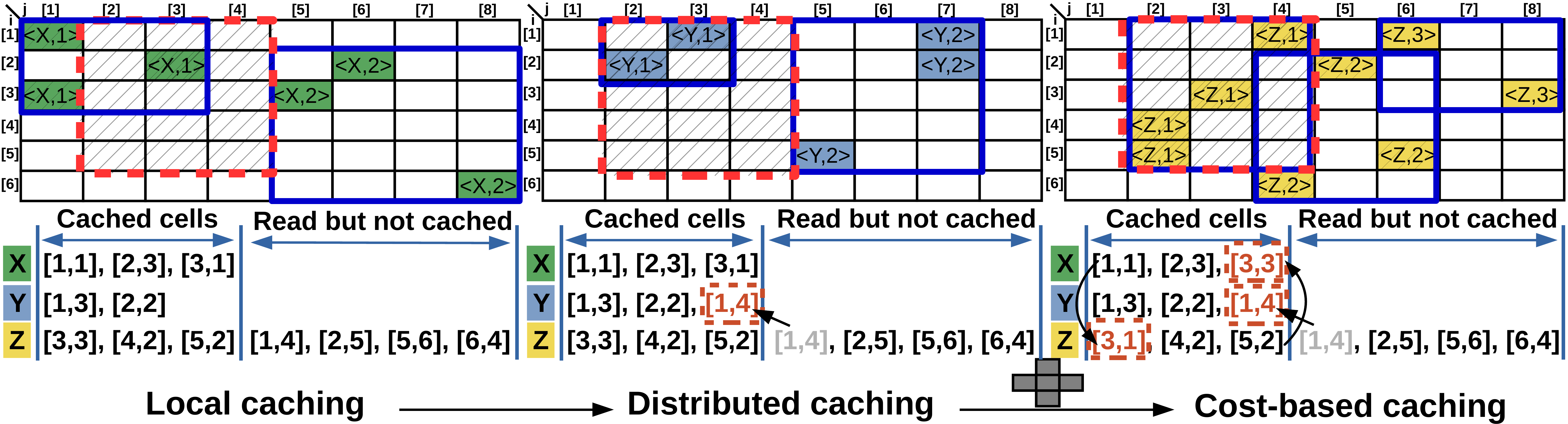}
\end{center}
\caption{High-level view of raw array distributed caching. The chunks in Figure~\ref{fig:array-example} are shown together with the corresponding bounding box separately for each server. The queried subarray is depicted as a dashed rectangle.}
\label{fig:cache-example}
\end{figure*}

\subsection{Cost-Based Chunk Caching}\label{ssec:chunk-caching}

Given the chunks extracted from the raw array and a memory caching budget at each server, the goal is to find an optimal caching plan to execute the query workload. In a centralized setting, cache optimality is defined as minimizing the number of cache misses, i.e., disk misses. In a distributed setting, another type of cache misses are present---chunks that have to be transferred over the network, i.e., network misses. The impact of these two measures depends on the relative throughput of the disk and the network. In the case of raw arrays, however, disk access always requires a complete file scan because chunks are not organized. Intuitively, this suggests that cache optimality should be defined exclusively in terms of disk misses. While this is true for queries that consider the chunks independently, in the case of array join queries, network misses are important because of chunk co-locality---chunk pairs that have to be processed together.

We illustrate the issues in array chunk caching on the example in Figure~\ref{fig:cache-example}. We set the memory budget on each node to at most $3$ cells. A self-similarity join query with a cross shape, i.e., $L^{1}(1)$, is executed over the dashed rectangle subarray. The example starts with a cold cache---the cache is empty in the beginning. The cache coordinator determines that chunks $f_{1,1}$, $f_{2,1}$, $f_{3,1}$, and $f_{3,2}$ overlap with the queried subarray. Since chunk $f_{3,2}$ does not have any cells that are contained by the query subarray, it is not considered for caching---it is split into two smaller chunks based on Algorithm~\ref{alg:evolving}. The other $3$ chunks overlap with the query, thus they are cached. In local caching, a chunk is cached on the node where it is stored. This is suboptimal for chunk $f_{3,1}$ which has $4$ cells---it cannot be cached entirely. In distributed caching, the memory budget is aggregated. As a result, chunk $f_{3,1}$ is cached completely by using the available entry at node $Y$. This eliminates the expensive disk misses incurred by partially cached chunks. In order to retrieve the uncached cell [1,4], a full scan over the file containing chunk $f_{3,1}$ is required. In order to execute the self-similarity join query, cell pairs ([1,3],[1,4]), ([1,3],[2,3]), ([2,2],[2,3]), ([2,3],[3,3]), and ([4,2],[5,2]) have to be resident on the same node. Since distributed caching does not consider network misses, only cell pairs ([1,3],[1,4]) and ([4,2],[5,2]) are collocated. A better caching configuration that collocates $3$ cell pairs is shown in Figure~\ref{fig:cache-example}.

We design a cost-based caching algorithm that weighs the chunks based on how they are accessed in the query workload. There are two components to the cost---access frequency and co-locality. Access frequency prioritizes chunks that are accessed in the near past. While this is similar to standard LRU, access is biased towards chunks belonging to the same raw file---a cache miss inside a file requires a complete scan of the file. The access frequency cost is included in the cache eviction section of the algorithm. Co-locality assigns chunk pairs that are joined together to the same node in the cluster. This cost is based on the query workload and is also biased towards more recent queries. The co-locality cost is included in the chunk placement section of the algorithm. The separation of cost-based caching into two sections is necessary because of the interaction between query processing and caching. If we consider them together, it is not clear how to define -- not to mention solve -- a cost formula that combines access and frequency.

\subsection{Cache Eviction}\label{ssec:distributed-caching:replacement}

After we generate the currently queried chunks, we have to identify chunks to evict from the cache---when the cache is full. Notice that only the chunks that overlap with the current query are considered for caching---chunks created because of non-overlapping splits are discarded. The important observation in caching raw array chunks is that we have to scan a file entirely even if only one accessed chunk is not cached. Therefore, we aim to cache all the queried chunks in a file. Conversely, if we evict a chunk from a file, we should evict all the other chunks from that file---independent of when they have been accessed. Since chunk-level LRU does not consider this correlation between chunks, it is likely suboptimal. File-level LRU -- on the other hand -- uses ineffectively the cache budget by caching non-accessed chunks. We include experiments for both of these two alternatives in Section~\ref{sec:experiments}.

We propose a novel cost-based cache eviction algorithm that integrates access with the raw file read savings for each unit price we pay, i.e., the additional cache budget we use. Specifically, the cost of a cached chunk is defined as:
\begin{equation*}
	\textit{cost}_{\textit{evict}}(Q_{l},f_{i},\{C_{j}\}) = w_{Q_{l}} \cdot \frac{\textit{size}(f_{i})}{\sum{\textit{size}(\textit{uncached } C_{j})}}
\end{equation*}
where $\{C_{j}\}$ is the set of chunks from file $f_{i}$ accessed by query $Q_{l}$. The weight of a query $w_{Q}$ is defined as an exponential function of the index $l$ of the query in the workload, e.g., $w(1)=2^{1},\dots,w(l)=2^{l}$. Essentially, the weight generates an exponential decay for the importance of the queries in the workload. The ratio between the file size and the size of the uncached chunks is larger when the number of chunks in the file required by a given query is smaller. If all the queried chunks are cached -- the case for the current query -- the ratio is infinity, thus the file is not evicted. 

Based on this eviction cost, we design an efficient greedy heuristic (Algorithm~\ref{alg:cache-replacement}) that evicts the chunks with the lowest cost. The algorithm takes as input the cache state, the chunks accessed by the current query, and the cache budget. It outputs the updated cache state that includes the newly accessed chunks. Instead of incrementally evicting chunks with the lowest cost, the heuristic selects the chunks with the highest cost to keep from the original cache state. This is necessary because of the correlation between chunks across queries. For example, a chunk with a low cost in a query may be evicted even though it is also accessed by an important query. If we decide to include the chunk, the decision is final. Moreover, this decision has an impact on other chunks. Thus, we increase the cost of all the correlated chunks to boost their chance of being cached (line~\ref{alg:evict:increase}). This is realized by evaluating the eviction cost dynamically without materialization.

\begin{algorithm}[htbp]
\caption{Cache Eviction}
\label{alg:cache-replacement}
\algsetup{linenodelimiter=.}

\begin{algorithmic}[1]

\REQUIRE Cache state as set of triples $S=\{(Q_{l},f_{i},\{C_{j}\})\}$ consisting of the chunks $j$ accessed from file $i$ at query $l$; Set of pairs $(f_{i},\{C_{j}\})$ consisting of the chunks $j$ accessed from file $i$ at the current query $Q_{l+1}$; Cumulated cache budget $B=\sum_{k}{B_{k}}$ consisting of local budgets $B_{k}$ at each node
\ENSURE Updated cache state $S'=\{(Q_{l+1},f_{i},\{C_{j}\})\}$

\STATE $S' \leftarrow \{(Q_{l+1},f_{i},\{C_{j}\})\}$
\WHILE{$\textit{budget}(S') < B$}
	\STATE Extract triple $t=(Q_{l},f_{i},\{C_{j}\})$ from $S$ such that $\textit{budget}(t\cup S') \leq B$ and $\textit{cost}(t)$ is maximum
	\STATE $S' \leftarrow S' \cup t$
	\STATE $S \leftarrow S \setminus t$
	\STATE Increase cost of triples $t'\in S$ that contain chunks in $t$\label{alg:evict:increase}
\ENDWHILE

\end{algorithmic}

\end{algorithm}

The time complexity of Algorithm~\ref{alg:cache-replacement} is proportional to the number of chunks in the cache state summed-up with the number of chunks accessed by the current query. If the total number of chunks -- cached and accessed -- is $N$, the complexity is $\mathcal{O}(N\log{N})$. The most time consuming operation is finding the best chunk to cache. This can be done by using a max binary heap having the eviction cost as key. Binary heap supports an efficient key increase operation required in line~\ref{alg:evict:increase} of the algorithm. Compared to LRU, the proposed eviction algorithm also considers all the cached chunks. Since the memory budget is bounded, the number of cached chunks cannot grow dramatically. Moreover, we can control the number of chunks by the threshold on the minimum number of cells inside a chunk (Algorithm~\ref{alg:evolving}).

\subsection{Cache Placement}\label{ssec:distributed-caching:placement}

Given the chunks selected to be cached, the next decision to make is how to allocate them over the distributed cache budget at the nodes. The main idea is to piggyback on the replication induced by query execution. Specifically, chunk-based join processing requires the chunk pairs to join to be collocated on the same node. Since one chunk joins with several other chunks, this induces replication. Notice that complete reshuffling is not desirable for sort-based partitioned arrays. The cache placement algorithm is performed after query execution and takes as input the location of all the currently cached chunks---including their replicas. The goal is to preserve a single copy of every chunk to better utilize the cache. Thus, we have to decide the node where to cache every chunk. We propose a solution that maximizes the co-locality of correlated chunks based on the entire query workload in order to reduce network traffic during query processing. Moreover, the communication cost of moving cached chunks across nodes is also considered.

To this end, we design a local search greedy algorithm for cache placement (Algorithm~\ref{alg:cache-placement}) derived from incremental array view maintenance~\cite{Zhao:IVM}. The algorithm takes as input the correlated chunk pairs across the performed queries, the location of the cached chunks, and the cache budget at every node. It outputs the updated cached chunk locations determined based on a cost function that measures the number of correlated collocated chunk join pairs:
\begin{equation*}
	\textit{cost}_{\textit{placement}}(C_{i},n,P',W) = \sum_{Q\in W}{w_{Q} \cdot \left|C_{j} \in P'_{n} \wedge (C_{i},C_{j}) \in Q\right|}
\end{equation*}
In this equation, $C_{i}$ is the current chunk to place, $n$ is a candidate node -- typically chosen from where a replica of $C_{i}$ exists -- $P'$ is the location of already placed chunks, and $W$ is the query workload. In general, the larger the number of collocated chunks, the larger the cost is. However, the contribution of a chunk pair is scaled by an exponentially decayed query weight that gives priority to recent queries. The order in which chunks are considered has an important impact on the algorithm. Our strategy is to take the chunks in increasing order of their number of replicas. Chunks with more replicas have more choices without network transfer when the local cache budget decreases.

\begin{algorithm}[htbp]
\caption{Cache Placement}
\label{alg:cache-placement}
\algsetup{linenodelimiter=.}

\begin{algorithmic}[1]

\REQUIRE Set $W=\{(Q_{l},\{(C_{i},C_{j})\})\}$ consisting of chunk pairs $(i,j)$ that join at query $l$; Set of locations $P=\{(C_{i},\{N_{k}\})\}$ specifying all the nodes $k$ that have a copy of cached chunk $i$ at current query $Q_{l+1}$; Cache budget $B_{k}$ at node $k$
\ENSURE Updated locations $P'=\{(C_{i},N_{k})\}$

\STATE $P' \leftarrow \{p \in P\}$, where $p$ has no replicas
\FOR{\textbf{each} $p=(C_{i},\{N_{k}\}) \in P$ with multiple replicas}
	\STATE Select node $n\in \{N_{k}\}$ such that $\textit{budget}_{n}(C_{i}) \leq B_{n}$ and $\textit{cost}(C_{i},n,P',W)$ is maximum
	\STATE $P' \leftarrow P' \cup (C_{i},n)$
\ENDFOR

\end{algorithmic}

\end{algorithm}

The complexity of Algorithm~\ref{alg:cache-placement} is $\mathcal{O}(|P|\cdot N + |W|)$, where $|P|$ is the number of cached chunks, $N$ is the number of nodes, and $|W|$ is the size of the query workload. Since $N$ is a small constant and $|P|$ can be bounded during the chunking process, the first term cannot become too large. $|W|$ can also be bounded by considering a limited number of previous queries. Moreover, the weight $w$ of ``old'' queries is small enough to be ignored due to exponential decay.


\section{Experimental Evaluation}\label{sec:experiments}

The objective of the experimental evaluation is to investigate the overall query processing performance and the overhead of cost-based caching for two types of array similarity join query workloads over the PTF catalog stored in two different file formats. These are real queries executed in the PTF pipeline for transient detection. We use the CSV LinkedGeoData dataset\footnote{\url{http://linkedgeodata.org}} in order to confirm the behavior of array caching on another file format. Specifically, the experiments are targeted to answer the following questions:

\begin{itemize}[leftmargin=*]
\item Does the proposed cost-based distributed caching improve upon file- and chunk-level LRU for a query workload?
\item How sensitive is cost-based caching to query patterns, cache budgets, and file formats?
\item How effective is query-driven array chunking?
\item Does the cache assignment improve the query communication time across a series of queries?
\item What is the execution time of the cost-based caching heuristics?
\end{itemize}

\subsection{Setup}

\textbf{Implementation.}
We implement caching for raw arrays as a \texttt{C++11} distributed multi-thread prototype that supports executing the array similarity join operator proposed in~\cite{Zhao:array-sim-join}. The catalog is stored at the coordinator. The distributed caching heuristic is also executed at the coordinator. It takes the input query from the query engine and generates the data communication plan containing information on chunk transfer and cache placement as output to the query engine. The query engine transfers chunks among nodes according to the data communication plan. The similarity join operator runs as a server on each node in the cluster. It manages a pool of worker threads equal to the number of CPU cores on each node. A worker thread is invoked with a pair of chunks that have to be joined. Chunks are retrieved directly from memory since the cache coordinator has already instructed each node to load the relevant chunks. This happens concurrently across all the workers. We use the CFITSIO\footnote{\url{https://heasarc.gsfc.nasa.gov/fitsio/}} and HDF5 libraries to read FITS and HDF5 files, respectively. Data serialization for network communication is optimized with the ProtoBuffers\footnote{\url{https://developers.google.com/protocol-buffers/}} library.

\textbf{System.}
We execute the experiments on a $9$-node cluster. The coordinator runs on one node while the other 8 nodes are workers. Each node has 2 AMD Opteron 6128 series 8-core processors (64 bit) -- 16 cores -- 28 GB of memory, and 4 TB of HDD storage. The number of worker threads is set to 16---the number of cores. Ubuntu 14.04.5 SMP $64$-bit with Linux kernel 3.13.0-43 is the operating system. The nodes are mounted inside the same rack and are inter-connected through a Gigabit Ethernet switch. The measured network bandwidth on a link is 125 MB/second. Since the disk bandwidth is in the same range, there is not a significant difference between the network and disk I/O.

\textbf{Data.}
We use the same two real datasets as in~\cite{Zhao:array-sim-join} for experiments. The \textit{PTF catalog} consists of 1 billion time-stamped objects represented in the equatorial coordinate system $(\textit{ra},\textit{dec})$. The range of the time coordinate spans over 153,064 distinct values, while for $\textit{ra}$ and $\textit{dec}$ we use ranges of 100,000 and 50,000, respectively. In array format, this corresponds to:\\
\hspace*{0.25cm}\texttt{PTF[time=1,153064;ra=1,100000;dec=1,50000]}\\
which is a sparse array with density less than $10^{-6}$. Objects are not uniformly distributed over this array. They are heavily skewed around the physical location of the acquiring telescope---latitude corresponds to $\textit{dec}$. Since HDF5 does not support sparse arrays natively, we store the PTF objects as an HDF5 table, i.e., relation. In FITS, data are stored as a binary table. Each tuple in HDF5 and FITS contains the dimensions and attributes for each cells in the original sparse array. The size of the PTF catalog is 343 GB in CSV, 262 GB in HDF5, and 221 GB in FITS. 

\textit{LinkedGeoData} stores geo-spatial data used in OpenStreetMap. We use the ``Place'' dataset which contains location information on roughly 3 million 2-D (long, lat) points-of-interest (POI). Since this is a too small dataset, we synthetically generate a larger dataset by adding 9,999 synthetic points with coordinates derived from each original point using a Gaussian distribution with $\mu = 0$ and $\sigma = 10$ miles~\cite{DasSarma:cluster-join}. In array format, this corresponds to:\\
\hspace*{0.25cm}\texttt{GEO[long=1,100000;lat=1,50000]}\\
With this replication, the size of GEO in CSV format is 1.6 TB. We split the entire dataset into 8,000 equal files.

\textbf{Query workloads.}
We extract $3$ query workloads from the original datasets using four methods---PTF-1, PTF-2, and GEO. Each workload contains $10$ queries. PTF-1 takes data exploration queries from the real workload that performs array similarity joins through all the detections on the $time$ dimension. For PTF-2, we simulate a typical data exploration pattern---shifting ranges and alternating queries. We extract $4$ range shifting queries from real workloads, enlarge the query ranges by a factor of 4 -- 2X on $ra$ $\times$ 2X on $dec$ -- and let them appear in the workload alternatively, e.g., $1,2,3,4,1,2,3,4,1,2$. GEO is a workload that contains queries with shifting ranges. The size of the query range is fixed. During query $1$ through $5$, the queried range shifts in the same direction---we increase latitude with a constant, e.g., $3,000$. Then we shift them back to the original position, e.g., $1,2,3,4,5,5,4,3,2,1$. Our small-size workloads mimic a data exploration behavior over raw arrays. In addition, the caching mechanism plays an even more important role when dealing with small workloads. Nonetheless, we also extract $100$ queries from the real PTF workload and perform a stress test of our caching algorithms.

\begin{figure*}[htbp]
\begin{center}
\includegraphics[width=\textwidth]{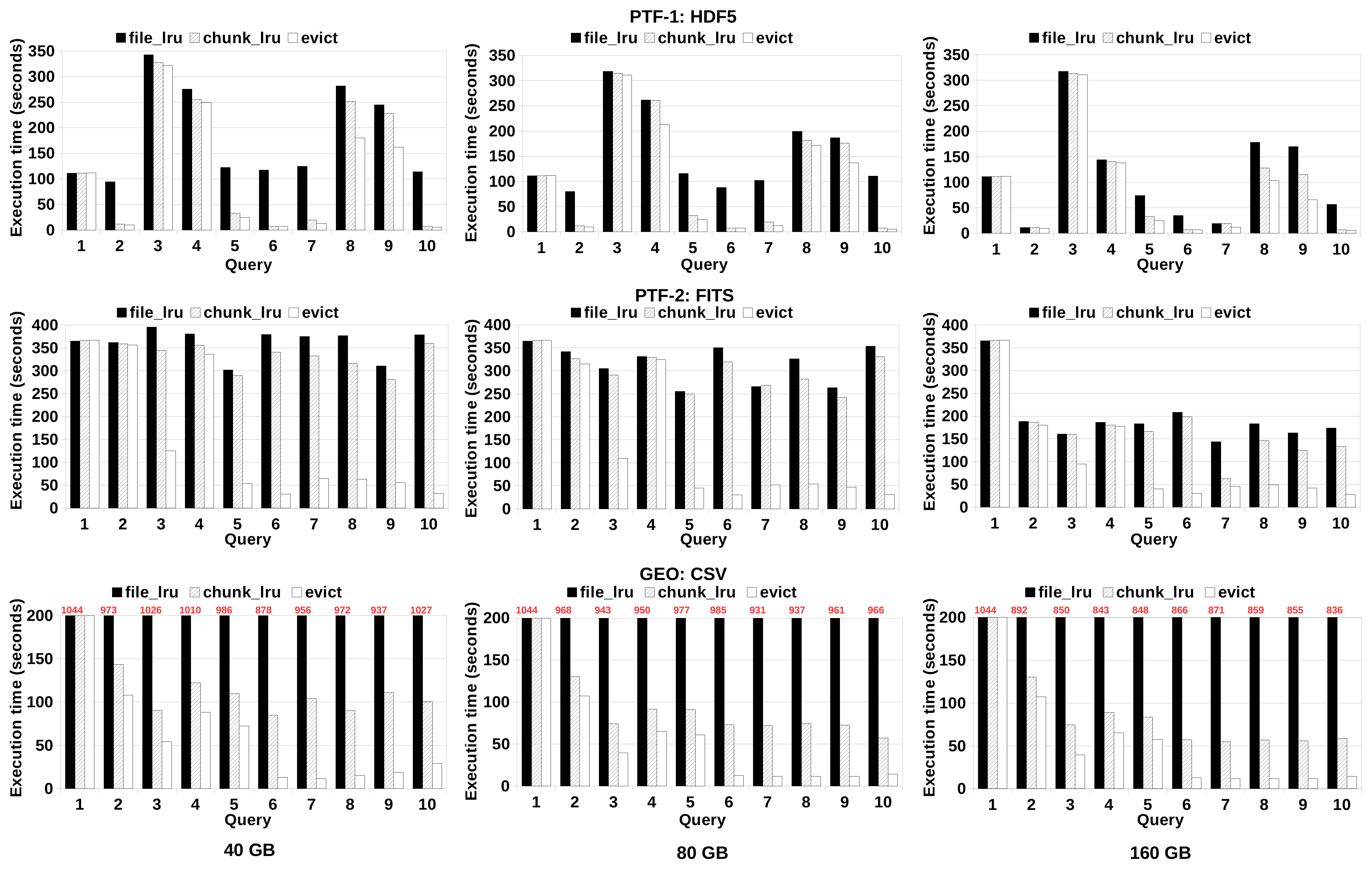}
\caption{Query execution time as a function of the cache memory budget. Three caching strategies -- file\_LRU, chunk\_LRU, and cost-based caching (evict) -- are compared across three workloads and three file formats. The time for cost-based caching includes raw file chunking, cache eviction, and cache placement. Chunk\_LRU includes only raw file chunking, while file\_LRU is the time for caching complete raw files using LRU.}
\label{fig:ptf-query}
\end{center}
\end{figure*}

\subsection{Results}

The code contains special functions to harness detailed profiling data. We perform all experiments at least 3 times and report the average value as the result. We always enforce data to be read from disk in the first query, i.e., cold caches.

\subsubsection{Execution Time}\label{ssec:experiments:data-preparation}

The execution time for caching is the time to read the raw files, extract queried data, and execute the query. The execution time is depicted in Figure~\ref{fig:ptf-query} for each individual query in the workload. We use file-level LRU -- denoted \textit{file\_lru} -- as the baseline. It utilizes all the memory across the cluster as a unified distributed memory and caches the least recent used files that contain queried data. \textit{file\_lru} works at file granularity because caching a subset of a file without the uncached bounding box still requires a full scan of the entire file---we have insufficient information to determine whether all the required data in the file are cached. \textit{chunk\_lru} computes bounding boxes during the query-driven chunking (Algorithm~\ref{alg:evolving}). It works at chunk level---caching the most recent queried chunks. \textit{evict} is the proposed cost-based caching that includes both cache eviction (Algorithm~\ref{alg:cache-replacement}) and cache placement (Algorithm~\ref{alg:cache-placement}) on top of \textit{chunk\_lru}. We use a cache budget of 40 GB, 80 GB, and 160 GB, respectively, to evaluate the scalability and behavior of the proposed algorithms. 40 GB is for the entire 8-node cluster, i.e., 5 GB for each node.

\textbf{PTF-1.}
The workload execution time for PTF-1 on HDF5 format exhibits large variations among queries. This is mostly due to the skewed nature of the dataset---some regions of the sky contain more detections than others. Some of the query execution times are very close to $0$ because raw file reading is not performed and the cache layout is not changed when all the queried data hit the cache. \textit{evict} always outperforms both LRU solutions. When we have 40 GB as cache budget, the difference varies across queries and is larger for the queries that share ranges---by as much as a factor of 20X for query 2, 6, 7, and 10. \textit{chunk\_lru} and \textit{evict} improve upon \textit{file\_lru} for all the queries, except the first---when they cannot even be applied. In the case of a larger cache budget, the difference between the LRU solutions and our cost-based caching is reduced because more available caching space makes the cache selection problem easier. In this case, \textit{evict} outperforms \textit{chunk\_lru} by 10-40\%.

\textbf{PTF-2.}
The execution time for PTF-2 on FITS format exhibits higher variance across the caching algorithms---especially when we have a small cache budget. The difference between \textit{file\_lru} and \textit{chunk\_lru} is smaller than for PTF-1 because the query ranges are enlarged and we only have a small budget. Since LRU always caches the most recent queried file, \textit{chunk\_lru} degenerates into \textit{file\_lru}. \textit{evict} outperforms \textit{chunk\_lru} starting at query 3 and becomes 5-10X faster after query 4. This is because \textit{evict} eliminates the case when large files are fully read only to extract a small amount of data. After increasing the cache budget to 160 GB, \textit{chunk\_lru} gets closer, however, \textit{evict} is still more than 3X faster on query 5, 6, and 10.

\begin{figure*}[htbp]
\begin{minipage}{.48\textwidth}
\centering
\includegraphics[width=.95\textwidth]{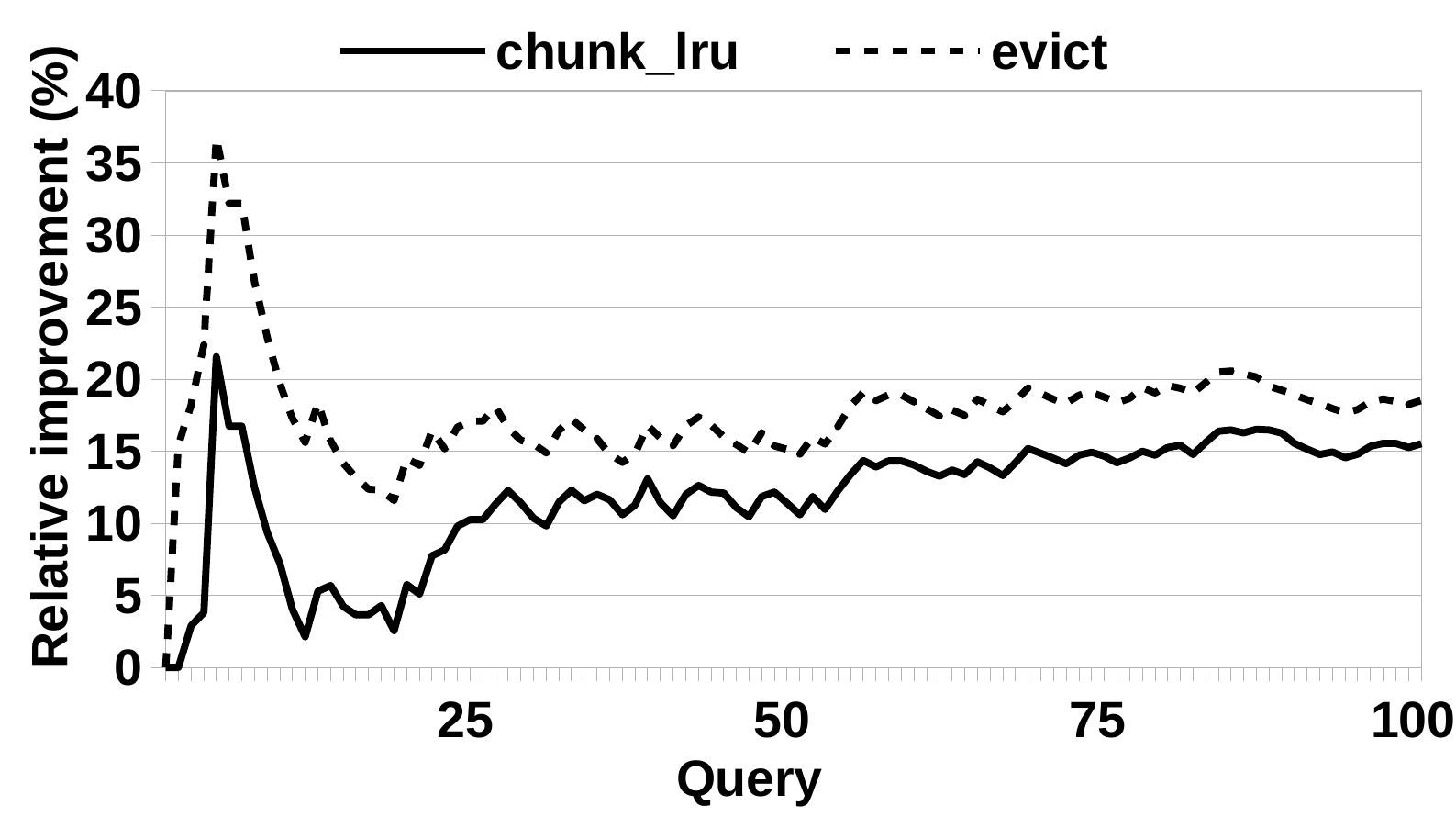}
\caption{Execution time improvement over \textit{file\_lru} on PTF in HDF5 format with 160 GB cache budget.}
\label{fig:100-query}
\end{minipage}
\hfill
\begin{minipage}{.48\textwidth}
\centering
\includegraphics[width=.95\textwidth]{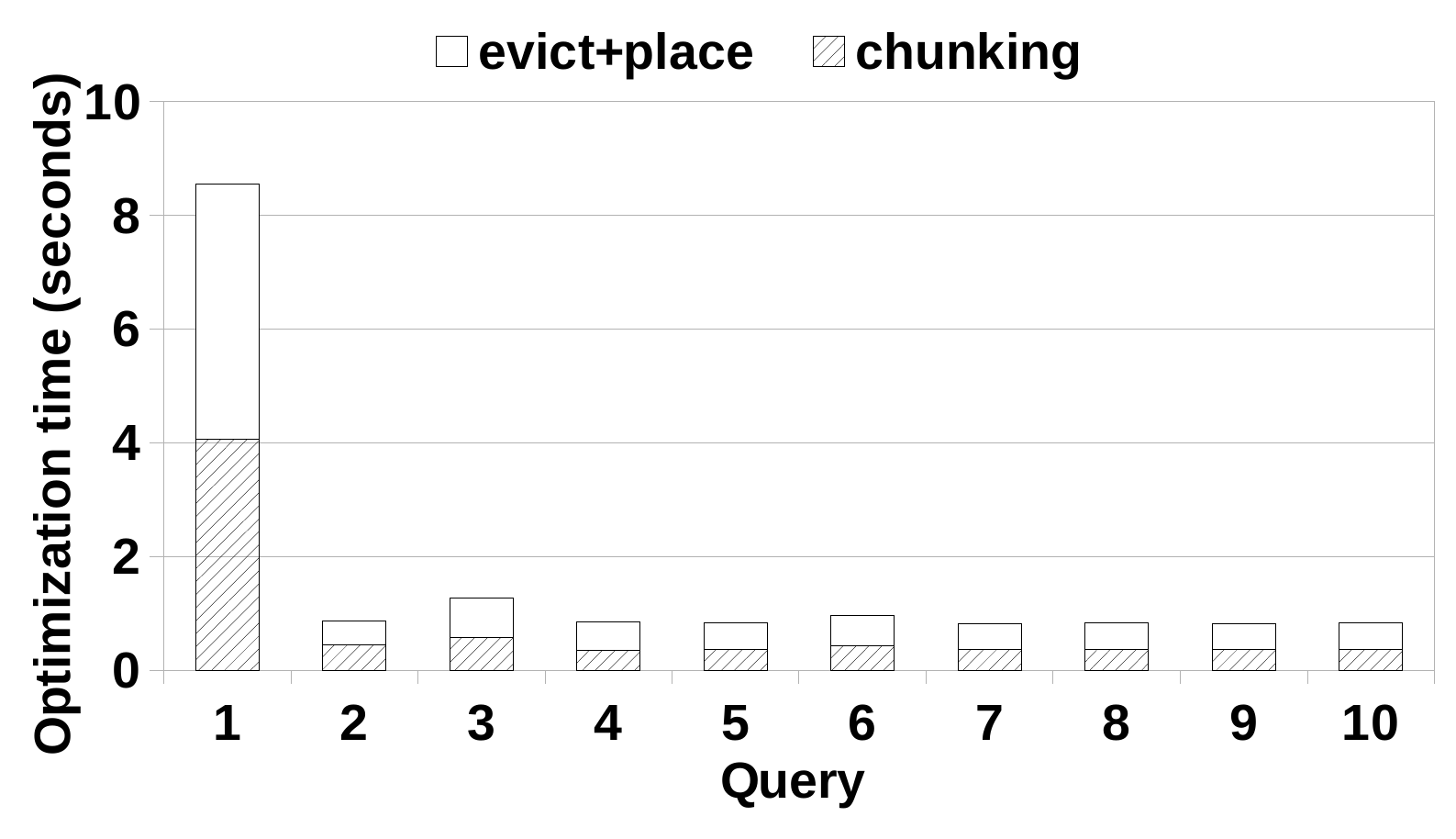}
\caption{Optimization time on GEO with a cache budget of 40 GB.}
\label{fig:opt-geo}
\end{minipage}
\end{figure*}

\textbf{GEO.}
Unlike the chunks in the real PTF dataset, chunking GEO files generates larger and sparser bounding boxes. \textit{file\_lru} cannot take advantage of the small cache budget. It is more than 10X slower than \textit{evict} which manages to decrease the chunk size whenever no overlaps are detected. Moreover, \textit{evict} outperforms \textit{chunk\_lru} by considering a weighted workload history to cover and read less files. This is exemplified by queries 6-10 which are the repetition of queries 1-5 in reverse order, where  \textit{evict} is 10X faster than \textit{chunk\_lru}.

\subsubsection{Scalability with the Number of Queries}\label{ssec:experiments:scalability}

Figure~\ref{fig:100-query} depicts the improvement in execution time generated by \textit{chunk\_lru} and cost-based caching over \textit{file\_lru} for every query in a workload of 100 PTF real queries. The cache budget is 160 GB, thus it favors the LRU algorithms. We observe that cost-based caching always outperforms the LRU algorithms by at most 20\%. This is a significant value for such a large cache budget and such a large number of queries since LRU has sufficient time to optimize its execution. Moreover, the chunks are dense enough to generate overlap with the queries.

\subsubsection{Optimization Time}\label{ssec:experiments:optimization}

The time to compute raw file chunking, eviction plan, and placement plan on LinkedGeoData with 40 GB cache budget is depicted in Figure~\ref{fig:opt-geo}. \textit{chunking} is the time used to generate and look-up query-driven chunking for raw files. \textit{evict+place} corresponds to the cost-based eviction and placement algorithms. Since the chunking is updated while we extract data from raw files, the updating time for splitting chunks is included in \textit{chunking}. For the first query, it takes 8 seconds to compute the original chunking and fill the cache budget. After that, all the optimizations finish within 1.5 seconds. The optimization time is acceptable considering the significant reduction in execution time it brings---as much as 800 seconds or more.

\subsubsection{Impact of Chunk Placement on Query Execution}\label{ssec:experiments:communication}

\begin{figure*}[htbp]
\begin{center}
\subfloat[]{\includegraphics[width=0.33\textwidth]{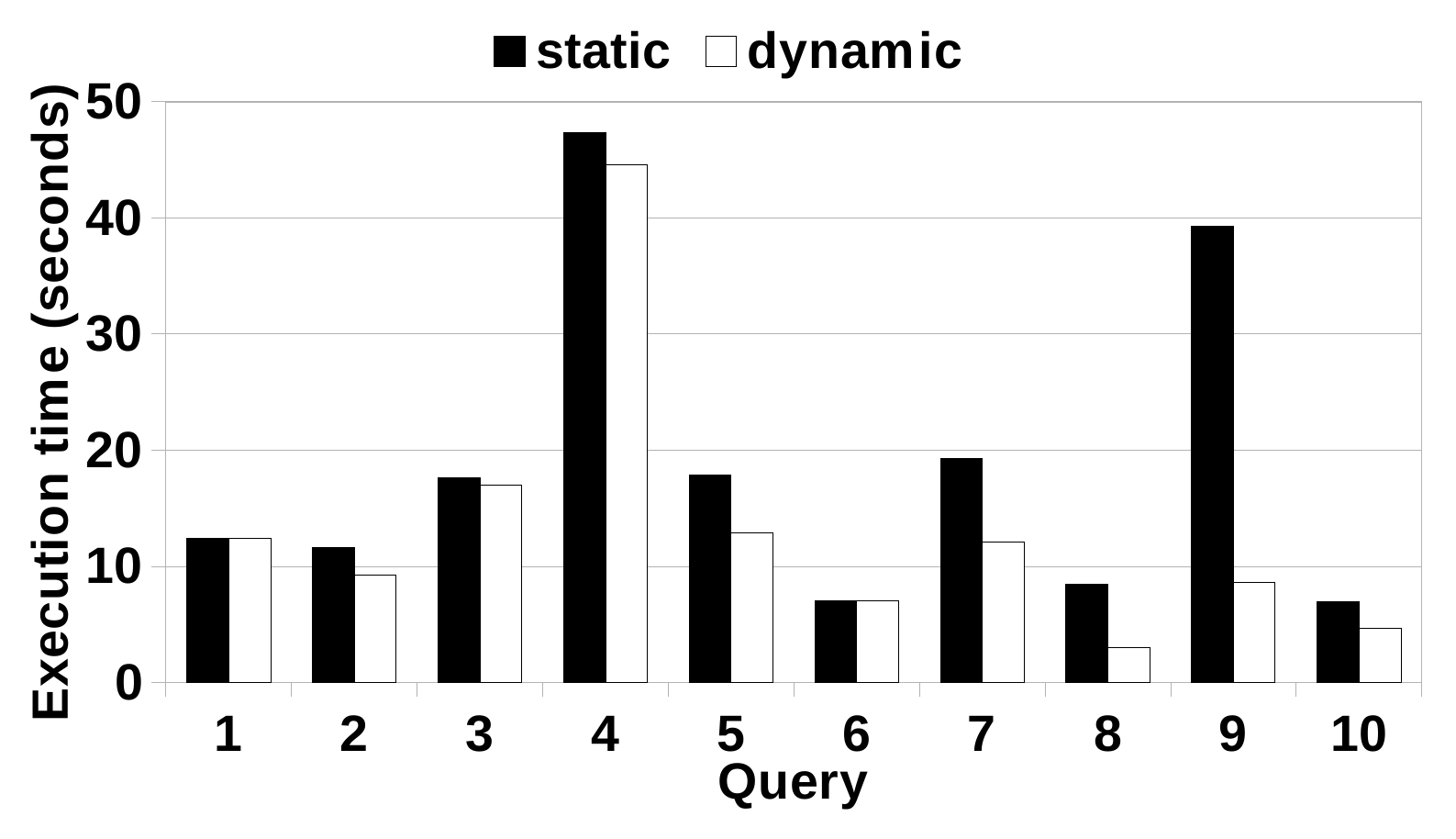}}\hfill
\subfloat[]{\includegraphics[width=0.33\textwidth]{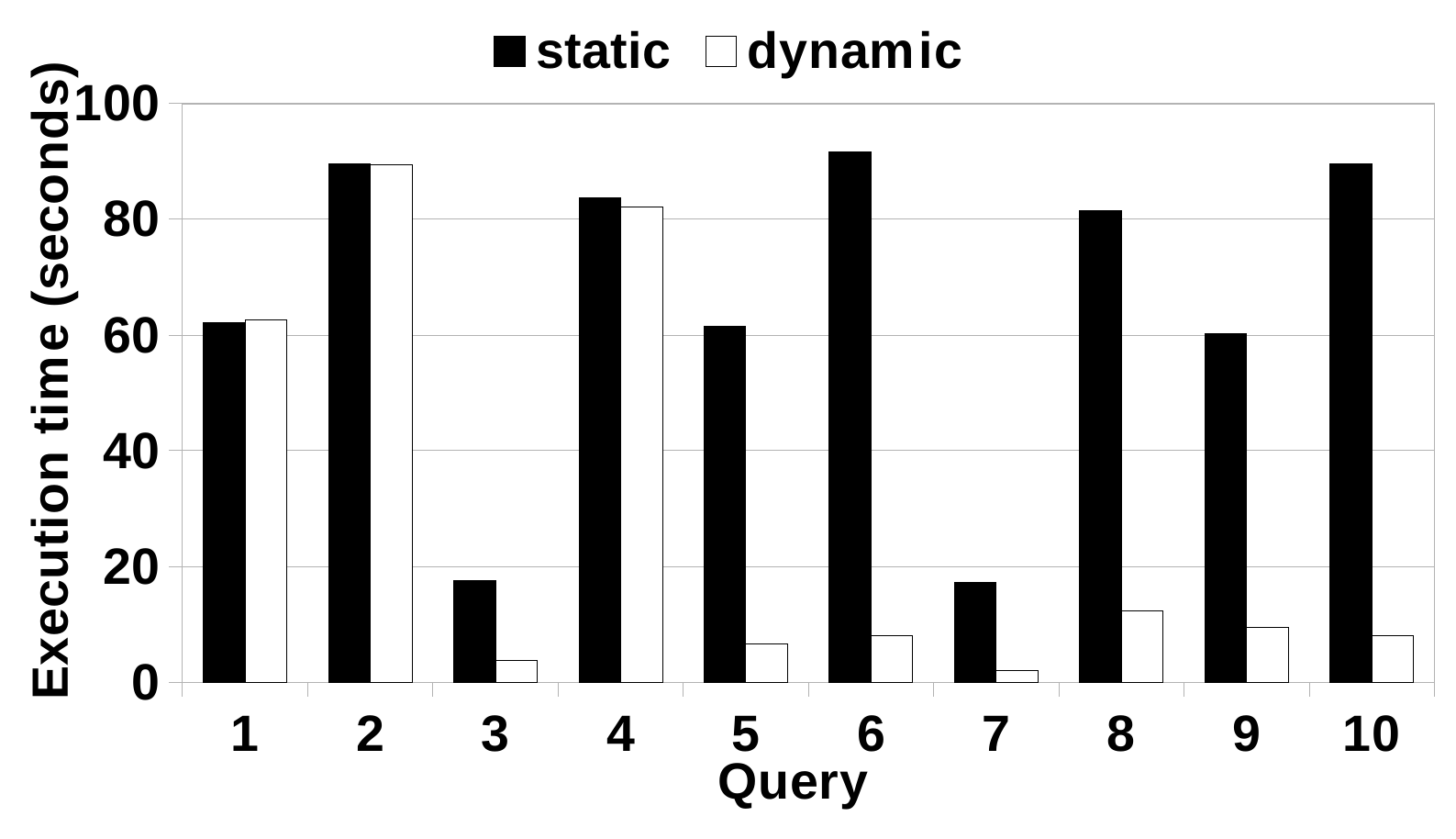}}\hfill
\subfloat[]{\includegraphics[width=0.33\textwidth]{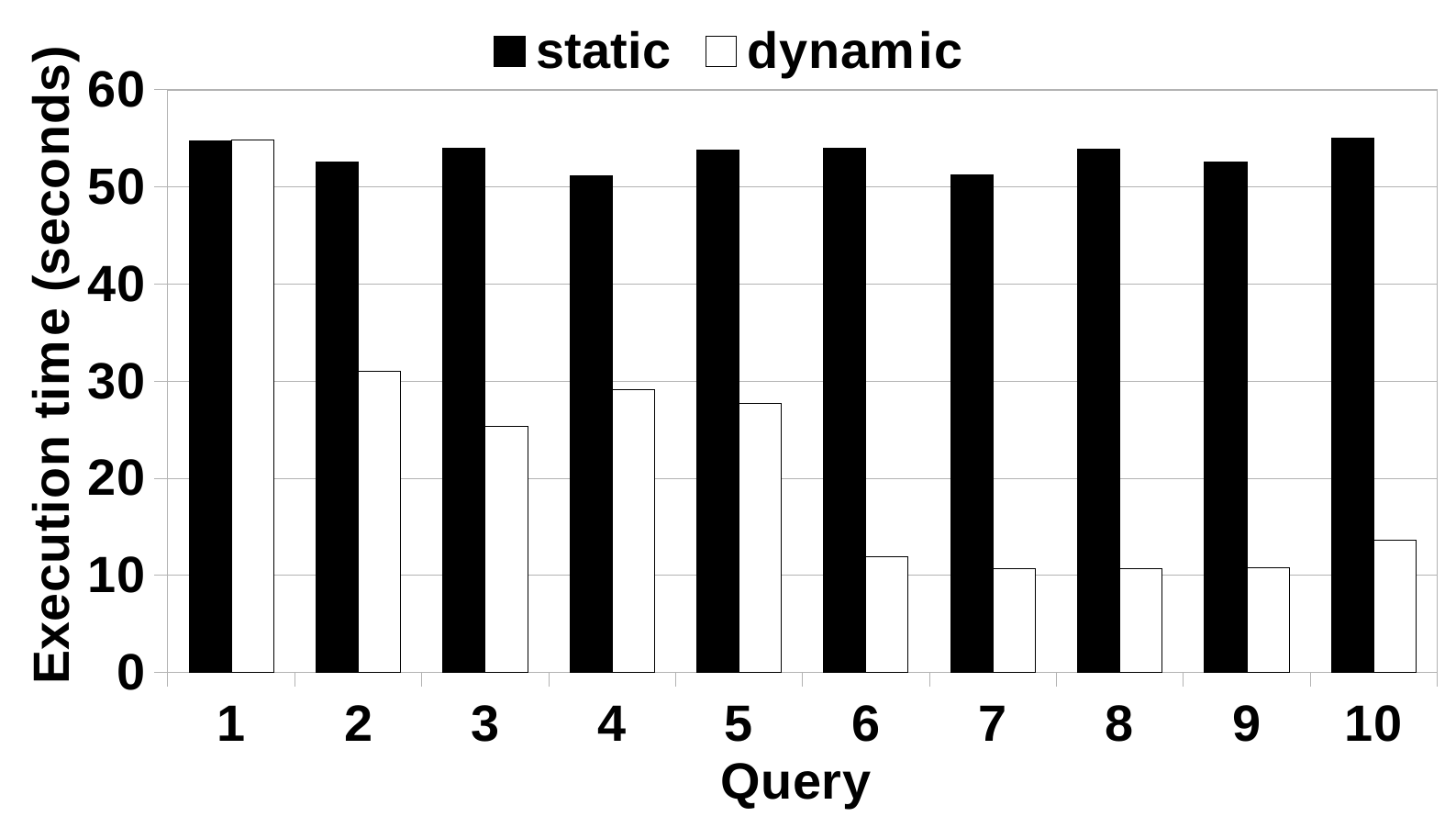}}\hfill
\caption{Reduction in similarity join execution time with cache placement (40 GB cache budget): (a) PTF-1, (b) PTF-2, and (c) GEO.}
\label{fig:comm}
\end{center}
\end{figure*}

Figure~\ref{fig:comm} shows the reduction in similarity join execution time due to cost-based cache placement. \textit{static} is the time without the cost-based cache placement, while \textit{dynamic} employs Algorithm~\ref{alg:cache-placement} to cache chunks considering query co-locality. Since we process queries in a multi-thread pipeline that overlaps CPU computation and network communication, the execution is I/O-bound with 16 threads enabled on each node---the network communication time is the bottleneck. Cache placement piggybacks on the join data transfer and reassigns cached chunks to other nodes. Thus, the placement algorithm improves the cache organization and alleviates the network I/O by assigning the data required to be joined on the same node. As expected, there is no difference for the first query because cache placement is not invoked in the beginning. For PTF-1 the effect of placement is not significant since there is no pattern in the workload. In contrast, for PTF-2 and GEO that have a shifting range query workload, \textit{dynamic} is 2-10X faster than the distributed LRU cache placement \textit{static}. This gap clearly proves the benefit of the placement algorithm.

\subsection{Discussion}\label{ssec:experiments:discussion}

The experimental results show that the proposed cost-based caching provides considerable improvement over distributed LRU for a query workload. Query-driven chunking improves over file-level LRU by 10-100X. On top of this improvement, the cost-based eviction algorithm accelerates caching by up to 100X. The file format has only a constant factor impact on data chunking---caused by the I/O format API libraries. When the cache budget is increased to 160 GB, the proposed algorithm is still more than 2X faster than chunk-level LRU. The cost-based cache placement algorithm improves the query communication time across a series of queries by a factor of 2-10X. The time taken by the optimizations is a small fraction of the data chunking time---more so when compared to the reduction it generates.

\section{Related Work}\label{sec:rel-work}

\textbf{Array databases.}
While many array databases have been proposed over the years, they implement caching following the standard parallel database approach. That is, LRU is executed locally at each node and only for the chunks assigned to the node. In the following, we focus only on how these systems handle caching and point the interested reader to~\cite{array-survey} for a comprehensive discussion on array database systems in general. RasDaMan~\cite{rasdaman} is a general middleware for array processing with chunks stored as BLOBs in a back-end relational database. RAM~\cite{ram} and SRAM~\cite{sram} provide support for array processing on top of the MonetDB~\cite{monetdb:overview} columnar database. They do not provide native support for arrays since arrays are represented as relations and array operations are mapped over relational algebra operators. While caching can be implemented in the middleware, this would replicate the functionality of the buffer manager in the back-end database system used to store the chunks. With integrated memory management and careful mapping from array chunks to disk pages, it is conceivable that an optimized array cache manager can be designed. RIOT~\cite{riot} is a prototype system for linear algebra operations over large vectors and matrices mapped into a standard relational database representation. Linear algebra operations are rewritten into SQL views and evaluated lazily. Caching is offloaded entirely to the underlying database. SciDB~\cite{scidb} is the most advanced shared-nothing parallel database system designed specifically for dense array processing. It supports multi-dimensional nested arrays with cells containing records, which in turn can contain multi-dimensional arrays. The buffer manager in SciDB is independent for each node and caches only local chunks. In order to support non-chunked sparse arrays, these have to be projected to a single dimension. During the loading process, SciDB performs repartitioning to create compact chunks. This process is extremely inefficient~\cite{extascid,Xing-Arraybridge}, though. SciHadoop~\cite{scihadoop} implements array processing on top of the popular Hadoop Map-Reduce framework which has only primitive caching support---the cache is replicated across all the nodes. ArrayStore~\cite{Soroush:arraystore}, TrajStore~\cite{trajstore}, and TileDB~\cite{tiledb} are storage managers optimized for multi-dimensional arrays and trajectories, respectively. They rely on the file system buffer manager for caching.

\textbf{Raw data processing.}
At a high level, we can group raw data processing into two categories. In the first category, we have extensions to traditional database systems that allow raw file processing inside the execution engine. Examples include external tables~\cite{oracle:external-tables,datastage,impala} and various optimizations that eliminate the requirement for scanning the entire file to answer the query~\cite{files-queries-results,nodb,data-vaults}. Modern database engines -- e.g., Oracle, MySQL, Impala -- provide external tables as a feature to directly query flat files using SQL without paying the upfront cost of loading the data into the system. NoDB~\cite{nodb} and Proteus~\cite{Ailamaki:proteus} enhance external tables by extracting only the attributes required in the query and caching them in memory for use in subsequent queries. Data vaults~\cite{data-vaults} and SDS/Q~\cite{sdsq} apply the same idea of query-driven just-in-time caching to scientific repositories. OLA-RAW~\cite{ola-raw} caches samples rather than full columns. Adaptive partial loading~\cite{files-queries-results} materializes the cached data in NoDB to secondary storage before query execution starts---loading is query-driven. ReCache~\cite{cost-based-caching} chooses the format in which to cache nested in-memory data adaptively. SCANRAW~\cite{scanraw} is a super-scalar adaptive external tables implementation that materializes data only when I/O resources are available. Instant loading~\cite{instant-loading} introduces vectorized SIMD implementations for tokenizing. RAW~\cite{raw} and its extensions VIDa~\cite{vida,Ailamaki:proteus} generate scan operators just-in-time for the underlying data file and the incoming query. The second category is organized around Hadoop MapReduce which processes natively raw data by including connector code in the Map and Reduce functions. Invisible loading~\cite{invisible-loading} focuses on eliminating the connector code by loading the already converted data inside a database. While similar to adaptive partial loading, instead of saving all the tuples into the database, only a fraction of tuples are stored for every query. None of these solutions supports in-situ processing over sparse arrays---the central contribution of this work. SciDB connectors for multi-dimensional arrays stored in HDF5 are introduced in~\cite{Xing-Arraybridge,Han-distributed-in-situ-array}. They allow SciDB to execute declarative array queries over dense arrays stored in HDF5 format by pushing down the subarray operator into the HDF5 read function. However, they do not support sparse arrays and caching---pushed to SciDB.

\textbf{Distributed caching.} 
In the context of relational databases and storage systems, there is extensive work on cache eviction algorithms such as the LRU-K~\cite{lruk}, DBMIN~\cite{chou1986evaluation}, and LRFU~\cite{lee2001lrfu}. Unlike our framework, these algorithms operate on fixed size pages -- not raw files of sparse arrays -- or for semantic caching~\cite{dar1996semantic}. In the web context, many caching policies have been developed for variable-size objects. Some of the most well-known algorithms in this space are LRU-Threshold~\cite{abrams1995caching}, Lowest-Latency-First~\cite{wooster1997proxy}, and Greedy-Dual-Size~\cite{cost-cache-algo}. In the context of Hadoop systems, Impala\footnote{\url{https://impala.apache.org/}} and Hortonworks\footnote{\url{https://hortonworks.com/}} allow users to manually pin HDFS files or partitions in the HDFS cache. This can be done automatically with adaptive algorithms in~\cite{Floratou-AdaptiveCaching}. In Spark\footnote{https://spark.apache.org/}, RDDs can be cached manually in Tachyon~\cite{Li-Tachyon}, a distributed in-memory file system. While these solutions work for raw files, they do not consider the multi-dimensional structure of arrays---they simply cache non-structured chunks. Distributed caching for spatial mobile data is considered in~\cite{Lubbe} and~\cite{proactive-cache}. They cache the chunks based on the spatial distance between chunks over the non-overlapping partition of the data---that is not applicable to the unorganized raw arrays.

\section{CONCLUSIONS AND FUTURE WORK}\label{sec:conclusions}

In this paper, we introduce a distributed framework for cost-based caching of multi-dimensional arrays in native format. The framework computes an effective caching plan in two stages. First, the plan identifies the cells to be cached locally from each of the input files by continuously refining an evolving R-tree index. In the second stage, a cost-based assignment of cells to nodes that collocates dependent cells in order to minimize the overall data transfer is determined. We provide a cost-based formulation for cache eviction that considers the historical query workload. A thorough experimental evaluation over two real datasets in three file formats confirms the superiority of the proposed framework over existing techniques in terms of caching overhead and workload execution time by as much as two orders of magnitude. In future work, we plan to explore how the proposed framework can be integrated with high-performance computing caching architectures such as the Burst Buffer\footnote{\url{www.nersc.gov/users/computational-systems/cori/burst-buffer}} and with the SciDB buffer manager.

\paragraph*{Acknowledgments}
This work is supported by a U.S. Department of Energy Early Career Award (DOE Career).

\bibliographystyle{abbrv}

\begin{thebibliography}{10}

\bibitem{tensorflow}
{M. Abadi et al.}
\newblock {TensorFlow: {A} System for Large-Scale Machine Learning}.
\newblock In {\em OSDI 2016}.

\bibitem{invisible-loading}
{A. Abouzied et al.}
\newblock {Invisible Loading: Access-Driven Data Transfer from Raw Files into
  Database Systems}.
\newblock In {\em {EDBT/ICDT 2013}}.

\bibitem{abrams1995caching}
M.~Abrams, C.~R. Standridge, G.~Abdulla, S.~Williams, and E.~A. Fox.
\newblock {Caching Proxies: Limitations and Potentials}.
\newblock 1995.

\bibitem{nodb}
I.~Alagiannis, R.~Borovica, M.~Branco, S.~Idreos, and A.~Ailamaki.
\newblock {NoDB: Efficient Query Execution on Raw Data Files}.
\newblock In {\em {SIGMOD 2012}}.

\bibitem{Altinel-CacheTables}
M.~Altinel, C.~Bornh{\"o}vd, S.~Krishnamurthy, C.~Mohan, H.~Pirahesh, and
  B.~Reinwald.
\newblock {Cache Tables: Paving the Way for an Adaptive Database Cache}.
\newblock In {\em VLDB 2003}.

\bibitem{datastage}
{N. Alur et al.}
\newblock {\em {IBM DataStage Data Flow and Job Design}}.
\newblock 2008.

\bibitem{cost-based-caching}
T.~Azim, M.~Karpathiotakis, and A.~Ailamaki.
\newblock {ReCache: Reactive Caching for Fast Analytics over Heterogeneous
  Data}.
\newblock {\em PVLDB}, 11(3), 2017.

\bibitem{rasdaman}
P.~Baumann, A.~Dehmel, P.~Furtado, R.~Ritsch, and N.~Widmann.
\newblock {The Multidimensional Database System RasDaMan}.
\newblock In {\em {SIGMOD 1998}}.

\bibitem{sdsq}
S.~Blanas, K.~Wu, S.~Byna, B.~Dong, and A.~Shoshani.
\newblock {Parallel Data Analysis Directly on Scientific File Formats}.
\newblock In {\em SIGMOD 2014}.

\bibitem{scidb}
{P. Brown et al.}
\newblock {Overview of SciDB: Large Scale Array Storage, Processing and
  Analysis}.
\newblock In {\em SIGMOD 2010}.

\bibitem{scihadoop}
J.~B. Buck, N.~Watkins, J.~LeFevre, K.~Ioannidou, C.~Maltzahn, N.~Polyzotis,
  and S.~Brandt.
\newblock {SciHadoop: Array-based Query Processing in Hadoop}.
\newblock In {\em {SC 2011}}.

\bibitem{cost-cache-algo}
P.~Cao and S.~Irani.
\newblock {Cost-Aware WWW Proxy Caching Algorithms}.
\newblock In {\em USENIX ITS 1997}.

\bibitem{scanraw}
Y.~Cheng and F.~Rusu.
\newblock {Parallel In-Situ Data Processing with Speculative Loading}.
\newblock In {\em {SIGMOD 2014}}.

\bibitem{extascid}
Y.~Cheng and F.~Rusu.
\newblock {Formal Representation of the SS-DB Benchmark and Experimental
  Evaluation in EXTASCID}.
\newblock {\em Distrib. and Parallel Databases}, 2014.

\bibitem{ola-raw}
Y.~Cheng, W.~Zhao, and F.~Rusu.
\newblock {Bi-Level Online Aggregation on Raw Data}.
\newblock In {\em SSDBM 2017}.

\bibitem{chou1986evaluation}
H.-T. Chou and D.~J. DeWitt.
\newblock {An Evaluation of Buffer Management Strategies for Relational
  Database Systems}.
\newblock {\em Algorithmica}, 1(1-4):311--336, 1986.

\bibitem{sram}
R.~Cornacchia, S.~H{\'e}man, M.~Zukowski, A.~P. de~Vries, and P.~Boncz.
\newblock {Flexible and Efficient IR using Array Databases}.
\newblock {\em VLDB Journal (VLDBJ)}, 17, 2008.

\bibitem{ssdb}
P.~Cudre-Mauroux, H.~Kimura, K.-T. Lim, J.~Rogers, S.~Madden, M.~Stonebraker,
  S.~B. Zdonik, and P.~G. Brown.
\newblock {SS-DB: A Standard Science DBMS Benchmark}.
\newblock \url{http://www.xldb.org/science-benchmark/}.

\bibitem{trajstore}
P.~Cudre-Mauroux, E.~Wu, and S.~Madden.
\newblock {TrajStore: An Adaptive Storage System for Very Large Trajectory Data
  Sets}.
\newblock In {\em {ICDE 2010}}.

\bibitem{dar1996semantic}
S.~Dar, M.~J. Franklin, B.~T. Jonsson, D.~Srivastava, M.~Tan, et~al.
\newblock {Semantic Data Caching and Replacement}.
\newblock In {\em VLDB 1996}.

\bibitem{DasSarma:cluster-join}
A.~D. Sarma, Y.~He, and S.~Chaudhuri.
\newblock {ClusterJoin: A Similarity Joins Framework using MapReduce}.
\newblock {\em PVLDB}, 7, 2014.

\bibitem{Duggan:array-joins}
{J. Duggan, O. Papaemmanouil et al.}
\newblock {Skew-Aware Join Optimization for Array Databases}.
\newblock In {\em {SIGMOD 2015}}.

\bibitem{Floratou-AdaptiveCaching}
A.~Floratou, N.~Megiddo, N.~Potti, F.~{\"O}zcan, U.~Kale, and
  J.~Schmitz-Hermes.
\newblock {Adaptive Caching in Big SQL using the HDFS Cache}.
\newblock In {\em SoCC 2016}.

\bibitem{furtado:tiling}
P.~Furtado and P.~Baumann.
\newblock {Storage of Multidimensional Arrays Based on Arbitrary Tiling}.
\newblock In {\em {ICDE 1999}}.

\bibitem{rtree}
A.~Guttman.
\newblock {R-trees: A Dynamic Index Structure for Spatial Searching}.
\newblock In {\em {SIGMOD 1984}}.

\bibitem{Han-distributed-in-situ-array}
D.~Han, Y.-M. Nam, and M.-S. Kim.
\newblock {A Distributed In-Situ Analysis Method for Large-Scale Scientific
  Data}.
\newblock In {\em BigComp 2017}.

\bibitem{proactive-cache}
H.~Hu, J.~Xu, W.~S. Wong, B.~Zheng, D.~L. Lee, and W.-C. Lee.
\newblock {Proactive Caching for Spatial Queries in Mobile Environments}.
\newblock In {\em ICDE 2005}.

\bibitem{monetdb:overview}
{S. Idreos et al.}
\newblock {MonetDB: Two Decades of Research in Column-Oriented Database
  Architectures}.
\newblock {\em IEEE Data Eng. Bull.}, 35(1), 2012.

\bibitem{files-queries-results}
{S. Idreos, I. Alagiannis et al.}
\newblock {Here Are My Data Files. Here Are My Queries. Where Are My Results?}
\newblock In {\em {CIDR 2011}}.

\bibitem{cost-cache-web}
S.~Irani.
\newblock {Page Replacement with Multi-Size Pages and Applications to Web
  Caching}.
\newblock In {\em STOC 1997}.

\bibitem{data-vaults}
M.~Ivanova, M.~L. Kersten, and S.~Manegold.
\newblock {Data Vaults: A Symbiosis between Database Technology and Scientific
  File Repositories}.
\newblock In {\em SSDBM 2012}.

\bibitem{Ailamaki:proteus}
M.~Karpathiotakis, I.~Alagiannis, and A.~Ailamaki.
\newblock {Fast Queries over Heterogeneous Data through Engine Customization}.
\newblock {\em PVLDB}, 9(12):972--983, 2016.

\bibitem{vida}
M.~Karpathiotakis, I.~Alagiannis, T.~Heinis, M.~Branco, and A.~Ailamaki.
\newblock {Just-In-Time Data Virtualization: Lightweight Data Management with
  ViDa}.
\newblock In {\em {CIDR 2015}}.

\bibitem{raw}
M.~Karpathiotakis, M.~Branco, I.~Alagiannis, and A.~Ailamaki.
\newblock {Adaptive Query Processing on RAW Data}.
\newblock {\em PVLDB}, 7, 2014.

\bibitem{impala}
{M. Kornacker et al.}
\newblock {Impala: A Modern, Open-Source SQL Engine for Hadoop}.
\newblock In {\em {CIDR 2015}}.

\bibitem{ptf:overview}
N.~M. Law et al.
\newblock {The Palomar Transient Factory: System Overview, Performance and
  First Results}.
\newblock {\em CoRR}, abs/0906.5350, 2009.

\bibitem{lee2001lrfu}
D.~Lee, J.~Choi, J.-H. Kim, S.~H. Noh, S.~L. Min, Y.~Cho, and C.~S. Kim.
\newblock {LRFU: A Spectrum of Policies that Subsumes the Least Recently Used
  and Least Frequently Used Policies}.
\newblock {\em IEEE Trans. on Computers}, 50(12):1352--1361, 2001.

\bibitem{Li-Tachyon}
H.~Li, A.~Ghodsi, M.~Zaharia, S.~Shenker, and I.~Stoica.
\newblock {Tachyon: Reliable, Memory Speed Storage for Cluster Computing
  Frameworks}.
\newblock In {\em SoCC 2014}.

\bibitem{aql:syntax}
K.-T. Lim, D.~Maier, J.~Becla, M.~Kersten, Y.~Zhang, and M.~Stonebraker.
\newblock {ArrayQL Syntax}.
\newblock
  \url{http://www.xldb.org/wp-content/uploads/2012/09/ArrayQL-Draft-4.pdf}.
\newblock [Online; February 2017].

\bibitem{Lubbe}
C.~L{\"u}bbe.
\newblock {Issues on Distributed Caching of Spatial Data}.
\newblock {\em University of Stuttgart, Germany, 2017}.

\bibitem{sgd-gpu}
Y.~Ma, F.~Rusu, and M.~Torres.
\newblock {Stochastic Gradient Descent on Highly-Parallel Architectures}.
\newblock {\em CoRR}, abs/1802.08800, 2018.

\bibitem{aql:algebra}
D.~Maier.
\newblock {ArrayQL Algebra: version 3}.
\newblock
  \url{http://www.xldb.org/wp-content/uploads/2012/09/ArrayQL_Algebra_v3+.pdf}.
\newblock [Online; February 2017].

\bibitem{instant-loading}
{T. Muhlbauer, W. Rodiger, R. Seilbeck et al.}
\newblock {Instant Loading for Main Memory Databases}.
\newblock {\em PVLDB}, 6(14), 2013.

\bibitem{GRAPPA}
J.~Nelson, B.~Holt, B.~Myers, P.~Briggs, L.~Ceze, S.~Kahan, and M.~Oskin.
\newblock {Latency-Tolerant Software Distributed Shared Memory}.
\newblock In {\em USENIX ATC 2015}.

\bibitem{Olma-Slalom}
M.~Olma, M.~Karpathiotakis, I.~Alagiannis, M.~Athanassoulis, and A.~Ailamaki.
\newblock {Slalom: Coasting through Raw Data via Adaptive Partitioning and
  Indexing}.
\newblock {\em PVLDB}, 10(10):1106--1117, 2017.

\bibitem{lruk}
E.~J. O'Neil, P.~E. O'Neil, and G.~Weikum.
\newblock {The LRU-K Page Replacement Algorithm for Database Disk Buffering}.
\newblock In {\em SIGMOD 1993}.

\bibitem{Ousterhout-RAMCloud}
J.~Ousterhout, P.~Agrawal, D.~Erickson, C.~Kozyrakis, J.~Leverich,
  D.~Mazi{\`e}res, S.~Mitra, A.~Narayanan, D.~Ongaro, and G.~Parulkar.
\newblock {The Case for RAMCloud}.
\newblock {\em Communications of the ACM}, 54(7):121--130, 2011.

\bibitem{tiledb}
S.~Papadopoulos, K.~Datta, S.~Madden, and T.~Mattson.
\newblock {The TileDB Array Data Storage Manager}.
\newblock {\em PVLDB}, 10(4):349--360, 2016.

\bibitem{array-survey}
F.~Rusu and Y.~Cheng.
\newblock {A Survey on Array Storage, Query Languages, and Systems}.
\newblock {\em CoRR}, abs/1302.0103, 2013.

\bibitem{Soroush:arraystore}
E.~Soroush, M.~Balazinska, and D.~Wang.
\newblock {ArrayStore: A Storage Manager for Complex Parallel Array
  Processing}.
\newblock In {\em {SIGMOD 2011}}.

\bibitem{mllib}
{E. Sparks et al.}
\newblock {MLI: An API for Distributed Machine Learning}.
\newblock In {\em {ICDM 2013}}.

\bibitem{cdn-caching}
V.~Sourlas, L.~Gkatzikis, P.~Flegkas, and L.~Tassiulas.
\newblock {Distributed Cache Management in Information-Centric Networks}.
\newblock {\em IEEE Trans. on Network and Service Management}, 10(3):286--299,
  2013.

\bibitem{sinew}
D.~Tahara, T.~Diamond, and D.~J. Abadi.
\newblock Sinew: A SQL System for Multi-Structured Data.
\newblock In {\em SIGMOD 2014}.

\bibitem{Heinis:thermal-join}
F.~Tauheedy, T.~Heinis, and A.~Ailamaki.
\newblock {THERMAL-JOIN: A Scalable Spatial Join for Dynamic Workloads}.
\newblock In {\em {SIGMOD 2015}}.

\bibitem{ram}
A.~R. van Ballegooij.
\newblock {RAM: A Multidimensional Array DBMS}.
\newblock In {\em {EDBT 2004}}.

\bibitem{wang2007workload}
X.~Wang, T.~Malik, R.~Burns, S.~Papadomanolakis, and A.~Ailamaki.
\newblock {A Workload-Driven Unit of Cache Replacement for Mid-Tier Database
  Caching}.
\newblock In {\em DASFAA 2007}.

\bibitem{oracle:external-tables}
A.~Witkowski, M.~Colgan, A.~Brumm, T.~Cruanes, and H.~Baer.
\newblock {Performant and Scalable Data Loading with Oracle Database 11g},
  2011.
\newblock {Oracle Corp.}

\bibitem{wooster1997proxy}
R.~P. Wooster and M.~Abrams.
\newblock {Proxy Caching that Estimates Page Load Delays}.
\newblock {\em Computer Networks and ISDN Systems}, 29(8):977--986, 1997.

\bibitem{Xing-Arraybridge}
H.~Xing, S.~Floratos, S.~Blanas, S.~Byna, K.~Wu, P.~Brown, et~al.
\newblock {ArrayBridge: Interweaving Declarative Array Processing with
  High-Performance Computing}.
\newblock {\em arXiv preprint arXiv:1702.08327}, 2017.

\bibitem{riot}
Y.~Zhang, H.~Herodotos, and J.~Yang.
\newblock {RIOT: I/O-Efficient Numerical Computing without SQL}.
\newblock In {\em {CIDR 2009}}.

\bibitem{SciQL-ideas}
Y.~Zhang, M.~Kersten, M.~Ivanova, and N.~Nes.
\newblock {SciQL: Bridging the Gap between Science and Relational DBMS}.
\newblock In {\em {IDEAS 2011}}.

\bibitem{vert-part-offline}
W.~Zhao, Y.~Cheng, and F.~Rusu.
\newblock {Vertical Partitioning for Query Processing over Raw Data}.
\newblock In {\em SSDBM 2015}.

\bibitem{Zhao:array-sim-join}
W.~Zhao, F.~Rusu, B.~Dong, and K.~Wu.
\newblock {Similarity Join over Array Data}.
\newblock In {\em {SIGMOD 2016}}.

\bibitem{Zhao:IVM}
W.~Zhao, F.~Rusu, B.~Dong, K.~Wu, and P.~Nugent.
\newblock {Incremental View Maintenance over Array Data}.
\newblock In {\em {SIGMOD 2017}}.

\end{thebibliography}

\end{document}